\documentclass[12pt]{article}

\newenvironment{wileykeywords}{\textsf{Keywords:}\hspace{\stretch{1}}}{\hspace{\stretch{1}}\rule{1ex}{1ex}}

\usepackage{amsmath,amssymb}
\usepackage{graphicx}
\usepackage{color}
\usepackage{dcolumn}
\usepackage{bm}
\usepackage[numbers,super,comma,sort&compress]{natbib}

\usepackage{subfigure}
\usepackage{multirow}
\usepackage{mathrsfs}
\usepackage{braket}
\usepackage{array}
\usepackage{rotating}

\newcommand{\bea}{\begin{eqnarray}}
\newcommand{\eea}{\end{eqnarray}}
\newcommand{\be}{\begin{equation}}
\newcommand{\ee}{\end{equation}}

\newcommand{\tj}[6]{ \begin{pmatrix}
		#1 & #2 & #3 \\
		#4 & #5 & #6
\end{pmatrix}}

\definecolor{background-color}{gray}{0.98}

\title{A complex Gaussian approach to molecular photoionization}

\author{Abdallah Ammar,
Lorenzo Ugo Ancarani,
Arnaud Leclerc\thanks{Laboratoire de Physique et Chimie Th\'eoriques,
Universit\'e de Lorraine, CNRS UMR 7019,  57070 Metz, France}}

\begin{document}

\maketitle

\begin{abstract}
We develop and implement a Gaussian approach to calculate partial cross-sections and asymmetry parameters for molecular photoionization.
Optimal sets of complex Gaussian-type orbitals (cGTOs) are first obtained by non-linear optimization, to best fit sets of Coulomb or distorted continuum wave functions for relevant orbital quantum numbers.
This allows us to represent the radial wavefunction for the outgoing electron with accurate cGTO expansions.
Within a time-independent partial wave approach, we show that all the necessary transition integrals become analytical, in both length and velocity gauges, thus facilitating the numerical evaluation of photoionization observables.
Illustrative results, presented for NH$_3$ and H$_2$O
 within a one-active-electron monocentric model, validate numerically the proposed strategy based on a complex Gaussian representation of continuum states.
\end{abstract}

\begin{wileykeywords}
complex Gaussian-type orbitals, continuum wavefunctions, photoionization, non-linear optimization, Gaussian integrals
\end{wileykeywords}

\clearpage


\begin{figure}[h]
\centering
\colorbox{background-color}{
\fbox{
\begin{minipage}{1.0\textwidth}
\begin{minipage}{0.5\textwidth}
\includegraphics[width=\textwidth]{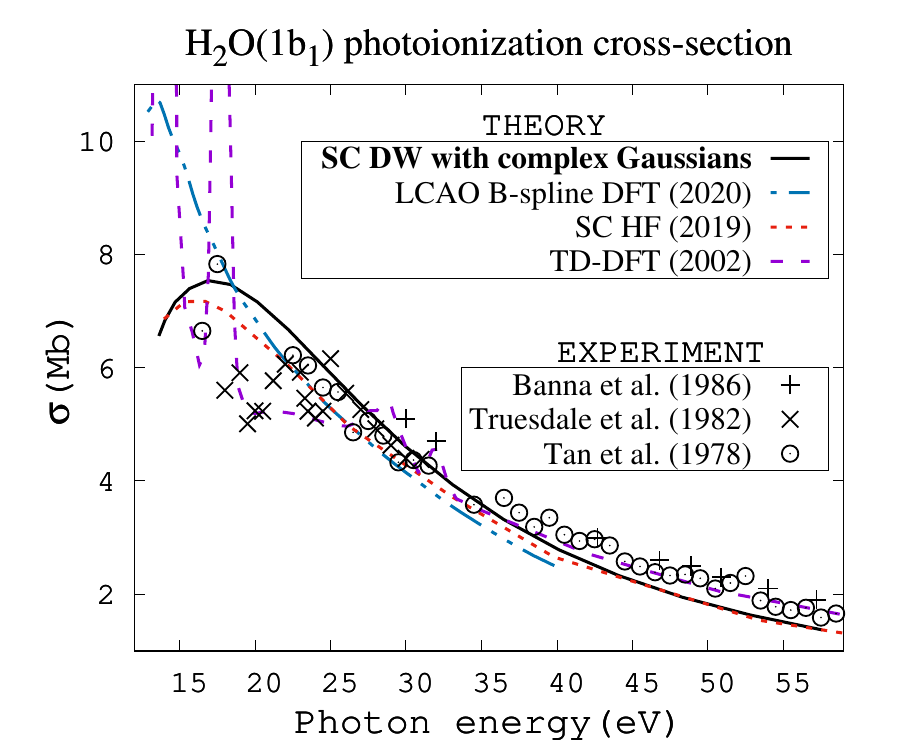}
\end{minipage}
\begin{minipage}{0.5\textwidth}
The theoretical description of molecular photoionization requires a good description of the outgoing electron wavefunction which oscillates up to infinite distances. In a single-center (SC) approach, the radial part of this continuum state can be efficiently represented up to a sufficiently large distance by an expansion on complex Gaussians (that is to say Gaussian functions with complex exponents). This greatly facilitates the calculation of cross-sections because the matrix elements can be evaluated analytically.
\end{minipage}
\end{minipage}
}}
\end{figure}

  \makeatletter
  \renewcommand\@biblabel[1]{#1.}
  \makeatother

\bibliographystyle{jcc}

\renewcommand{\baselinestretch}{1.5}
\normalsize

\clearpage

\section*{\sffamily \Large INTRODUCTION}

Ordinary Gaussian-type functions\cite{boys1950,hill2013} are not especially renowned for their adequacy to the theoretical description of continuum states. Those are involved, for example, in dynamical processes where an electron escapes from the system.
In a recent theoretical study about molecular photoionization,\cite{moitra2020} one can read that ``conventional Gaussian- or Slater-type orbitals are unsuitable for this purpose."
Everyone can intuitively agree with that, given that an outgoing electron possesses a wavefunction with an oscillating and non-decreasing behaviour, whereas Gaussian-type functions do not feature any nodes and always go to zero, more or less rapidly depending on the exponent values. Such properties
are intrinsically not compatible with the asymptotic behaviour of continuum states.
This said, if only there were a chance that some Gaussian set manages to attenuate these apparent incompatibilities, this would be very interesting from a numerical point of view in many applications involving electronic detachment processes. Many Gaussian integrals could then be expressed in closed form and, in the molecular case, multicentric integrals which naturally arise in cross-section calculations could be drastically simplified by using the Gaussian product theorem.\cite{shavitt1963,besalu2011}

In this paper we focus on photoionization,
which plays an important role in atomic and molecular physics and is strongly related to several other sub-domains of physics, including for example applications in astrophysics,\cite{bautista1998,bautista2001} 
plasma science \cite{levine1973,killian1999,foster2010} or
biology.\cite{hubbell1999,hubbell2006}
Photoionization has been studied since the early days of quantum mechanics, and the literature is quite vast.
Nowadays, while a theoretical-experimental agreement
can be generally observed for atoms, this is  not always the case for molecular targets,
especially in the near-threshold region where autoionizing resonant structures are difficult to describe,
as pointed out for example in refs.
\citenum{granados2016,novikovskiy2019}.
Such discrepancies are partly related to several theoretical difficulties coming from the multicentric nature of the target.
Recent photoelectron spectroscopy experiments on molecules such as those presented in refs. \citenum{stolow2008,suzuki2012,wolf2019}
provide an ongoing and renewed interest,  and a challenge for theoreticians.
Indeed, the process provides an indirect tool to test the theoretical description of the target before and after the interaction, and thus our capacity of describing correlation and
many-body effects for both bound and unbound electronic states.
A number of methods have been proposed in the past, and some are currently being developed with the aim of improving the accuracy of the dynamical parameters describing photoionization (see, for example, refs.
\citenum{moitra2020,gozem2015,novikovskiy2019}).
We are precisely interested in this difficult problem and we consider it as a good test case for a Gaussian representation of continuum states.

For bound states, the efficiency of Gaussian-type orbitals (GTOs) needs no further proof. For a review about the use of Gaussian basis sets in molecular calculations, see for example ref. \citenum{hill2013}.
However, in scattering processes,  as for example photoionization of neutral species or photodetachment of anions, an efficient description of continuum states is also required. What is meant by efficient is that the continuum wavefunction describing an incident or escaping charged particle, for example an outgoing electron, should be a good compromise between accuracy over a sufficiently large radial domain and numerical efficiency.
In terms of accuracy, neither standard GTOs (with real exponents) nor Slater-type orbitals seem very adapted to the task of describing oscillating, non-decreasing continuum wavefunctions, as already stated above.
However, as far as numerical efficiency is concerned, the use of Gaussian-type functions may lead to strong savings in terms of computational time.
Among these two observations, the first (pessimistic) one about the inadequacy to deal with the continuum seems to have outshined the second (optimistic) one on numerical gain; indeed,  in the literature there is a relatively low number of theoretical studies using Gaussian sets for describing continuum states,\cite{kaufmann1989,nestmann1990,faure2002,fiori2012}
sometimes within a hybrid basis set also including discrete variable representations\cite{rescigno2005}
or B-splines.\cite{marante2014,masin2014}

In a recent work,
we have developed an optimization method able to provide sets of complex Gaussians (\emph{i.e.} Gaussian-type orbitals with a complex exponent) able to reproduce accurately Coulomb waves,
and whose scattering applications in the benchmark case of atomic hydrogen demonstrated their reliability.\cite{ammar2020}
The complex Gaussian-type orbitals (cGTOs) are defined as
\be
G(r) = e^{- \alpha r^2} \; \text{ with }
\alpha \in \mathbb{C} ,
\label{eq:cgto}
\ee
where $\text{Re}(\alpha) > 0$ ensures that the functions are square integrable.
cGTOs were first introduced in the context of resonance stabilization calculations.\cite{mccurdy1978,mccurdy1982,isaacson1991,white2017}
More recently cGTOs were used in atomic photoionization calculations\cite{matsuzaki2016,matsuzaki2017} with an optimization method based on the Prony's algorithm used earlier by Huzinaga.\cite{huzinaga1965}
In ref. \citenum{ammar2020} we have focused our efforts on the numerical method used to perform cGTOs optimization in order to improve their numerical efficiency as a continuum representation.
cGTOs oscillate with a spatial frequency increasing with the radial coordinate $r$, a feature that is not physically satisfactory.
On the other hand, in comparison with real GTOs (rGTO), this oscillatory behavior due to the complex exponent leads to more stable expansions of physically sound continuum wavefunctions.
We have also shown that optimized cGTO sets can actually be used to calculate accurate ionization cross-sections for atomic hydrogen, under photon or electron impact.
It turns out that the necessary matrix elements can be evaluated analytically, making the approach numerically efficient.

Here we wish to adopt a similar strategy by presenting cGTO calculations of molecular photoionization differential cross-sections
and asymmetric parameters,
using a number of approximations:
we use a monocentric description of the targets, considering only small molecules of type AH$_n$ which are known to reasonably bear such an approximation;
we restrict ourselves to a one-active-electron description;
we assume the molecular targets to be randomly oriented;
we work in the dipolar approximation with linearly polarized light;
the initial bound orbitals of the target are described in terms of Slater-type orbitals with parameters taken from Moccia.\cite{moccia1964I,moccia1964II,moccia1964III}
The principal aim is not to develop a sophisticated chemical model or to get as close as possible to the available experimental results, but rather to prove that it is indeed possible to obtain converged results on differential photoionization cross-sections for molecules using pure cGTO expansions of the radial continuum wavefunction. Such a representation allows us to get closed-form expressions for the transition matrix integrals, the numerical evaluation of which becomes very fast.
The monocentric (also often named Single Center) approximation is an important restriction, and in a near future we may definitely consider similar calculations with a multicentric description of the target.
The main idea of the present paper is to show that theoretical photoionization cross-sections keep the same level of accuracy within a given model, even when cGTO expansions are used to describe the outgoing electron.

In the rest of the paper, we first summarize the photoionization model and parameters in section~\ref{sec_theory}. We introduce the fitting method that we have used to find optimal cGTO expansions of the (monocentric) continuum functions on a definite domain, adapted to molecular applications. Then, integrals needed for the photoionization cross-section calculations are detailed.
Finally, in section~\ref{sec_results} come the results on two illustrative examples,
the Ne-like molecules
 NH$_3$ and H$_2$O, followed by some concluding remarks.
Unless otherwise stated, we use Hartree atomic units throughout the paper.

\section{\sffamily \Large GAUSSIAN-SLATER MOLECULAR PHOTOIONIZATION MODEL \label{sec_theory}}

\subsection{\sffamily \large Photoionization parameters \label{subs_PIparameters}}


During the photoionization process, one electron of the target is ejected from an initial bound orbital $\phi_{i}(\mathbf{r})$ to a continuum state $\psi_{\mathbf{k_e}}^-(\mathbf{r})$ with wavevector $\mathbf{k_e}$ (kinetic energy $E_{k_e}=k_e^2/2$ and solid angle $\widehat{k_e}$).
We restrict ourselves to a simple one-active-electron model with electronic coordinate $\mathbf{r}$, all other electrons remaining unaffected.
We work in the dipole approximation and the photon of energy $E_{\gamma}$ is linearly polarized along a direction  $\hat{\epsilon}$.
The energy conservation reads
\begin{equation}
E_{\gamma} = E_{k_e} + V_{\text{ion}}
\text{,}
\end{equation}
where $V_{\text{ion}}$ is the ionization energy associated with the initial state.

The photoelectron angular distribution
(sometimes quoted with the acronym PAD in the literature)
is given by the differential cross-section, which should be averaged over all the possible molecular orientations. In a time-independent framework, it is given by:\cite{bethe1957,drake1996}
\begin{equation}
\frac{d\sigma}{d\widehat{k_e}} =
 {\cal N}_i
 \frac{4 \pi^2 k_e E_{\gamma}}{c }
 \int \, d\widehat{\mathscr{R}}
\left| T_{i \mathbf{k_e}} (\widehat{\mathscr{R}})
 \right|^2
 \label{eq:PICSdiff}
\end{equation}
where $\widehat{\mathscr{R}}$ stands for the molecular orientation with respect to the laboratory frame (Euler angles),
${\cal N}_i$ denotes the initial number of electrons in the concerned molecular orbital
and $c$ is the speed of light.
Although not explicitly written, the cross-section \eqref{eq:PICSdiff} is also differential with respect to the photoelectron energy.
The key quantities to be calculated are the dipole transition moment elements $T_{i\mathbf{k_e}}$.
In length and velocity gauges, they are respectively defined by
\begin{equation}
\begin{aligned}
T_{i\mathbf{k_e}}^{(\mathscr{L})}
&=
\braket{\psi_{\mathbf{k_e}}^-| \hat{\epsilon} \cdot \mathbf{r} |\phi_{i}}
\label{eq:Tlength}
\end{aligned}
\end{equation}
or
\begin{equation}
T_{i\mathbf{k_e}}^{\mathscr{(V)}}
=  \frac{1}{\imath E_{\gamma}} \braket{\psi_{\mathbf{k_e}}^-| \hat{\epsilon} \cdot
\mathbf{p} |\phi_{i}}
\label{eq:Tvelocity}
\end{equation}
with $ \mathbf{p} = - i \nabla$.
It is worth reminding that, while calculated cross-sections should be in principle gauge-invariant,  using an approximate Hamiltonian or non-exact wavefunctions may lead to substantial differences between
length and velocity gauges results.\cite{drake1996}

In the case of a linearly polarized photon, the differential cross-section for randomly oriented molecules can also be expressed using an energy-dependent asymmetry parameter $\beta$, through:\cite{chandra1987,machado1990}
\begin{equation}
\frac{d\sigma}{d\widehat{k_e}}  =
\frac{\sigma(k_e)}{4 \pi} \left[ 1 + \beta P_2(\cos{(\theta)} \right]
\label{eq:asymmetry_parameter}
\end{equation}
where
\be
\sigma(k_e) = \int \frac{d\sigma}{d\widehat{k_e}} \, d\widehat{k_e}
\label{eq:totalCS}
\ee
is the integrated cross-section
(or partial ionization cross section,
sometimes quoted with the acronym PICS in the literature),   $P_2(x)=\frac{1}{2}(3x^2-1)$ is the second Legendre polynomial, and
 $\theta= \widehat{ (\hat{\epsilon} , {\bf k_e}) }$
 is the scattering angle in the laboratory frame.
Compared to the integrated cross-section, the asymmetry parameter $\beta$ regulates the photoelectron angular distribution
and contains thus more detailed information; as a consequence, it is also more sensitive to the quality of the involved wavefunctions.

A typical photoionization calculation thus needs sufficiently accurate wavefunctions for the initial and continuum states to ensure that integrals \eqref{eq:Tlength} or \eqref{eq:Tvelocity}
yield reasonable results.

\subsection{\sffamily \large Model wavefunctions \label{subs_wavefunctions}}

The initial molecular orbitals of the target are taken from Moccia (ref. \citenum{moccia1964II} for NH$_3$ and ref.  \citenum{moccia1964III} for H$_2$O),
who used a self-consistent-field calculation.
He proposed a monocentric expansion in terms of $N_i$ Slater-type orbitals with real coefficients and real spherical harmonics, centered on the heaviest atom. We transform the data to work with complex spherical harmonics
$Y_{l}^{m}(\hat{r})$
and complex coefficients
$C_{ij}$,
and use the following form,
\begin{equation}
\phi_i(\mathbf{r}) = \sum_{j=1}^{N_i}
C_{ij}
 r^{n_j-1} e^{-\zeta_j r}
Y_{l_j}^{m_j}(\hat{r}) \text{.}
\label{eq:Moccia_phi}
\end{equation}
with normalization $\braket{\phi_i|\phi_i} =1$.

For the outgoing electron, we shall use a standard single center
partial wave expansion,\cite{messiahbook}
\begin{equation}
\psi_{\mathbf{k_e}}^-(\mathbf{r}) = \sqrt{\frac{2}{\pi}}
\sum_{l,m} (\imath)^l
e^{-\imath \delta_l }
\frac{u_{l,k_e}(r)}{k_e r} Y_l^m(\hat{r}) Y_l^{m*}(\widehat{k_e})
\label{eq:psi_f}
\end{equation}
where  $\delta_l$ denotes the phase shift for a given angular momentum $l$.
The radial part of the wavefunction $u_{l,k_e}(r)$ satisfies the following equation,
\begin{equation}
\left[ -\frac{1}{2} \frac{d^2}{dr^2} + \frac{l(l+1)}{2r^2}
+ U^{\text{mol}}(r)  \right] u_{l,k_e}(r)
= \frac{k_e^2}{2} u_{l,k_e}(r)
\label{eq:fct_disto_DE}
\end{equation}
for  a central potential $U^{\text{mol}}(r)$. This one-dimensional differential equation can be numerically solved by using, for example, the RADIAL code.\cite{salvat2019}

As a very first approximation, we assume that the potential felt by the photoelectron is Coulombic, $U^{\text{mol}}(r) \simeq -z/r$, with a point charge $z=1$, so that the continuum wavefunction is a pure Coulomb wave.
In this case the phase shift
$\delta_l =\arg \left( \Gamma(l+1 + \imath \eta) \right)$
with the  Sommerfeld parameter $\eta = -z/k_e $,
and the radial functions are the (real) regular Coulomb functions
\begin{equation}
\begin{aligned}
u_{l,k_e}(r) = & F_l(\eta,k_e r)  \\
=  &
(2k_e r)^{l+1} e^{-\frac{\pi \eta}{2}}
\frac{\left|\Gamma\left(l+1+\imath \eta \right)\right|}
{2\Gamma\left(2l+2 \right)} e^{\imath k_e r}
\\
& \times \,
\mathstrut_1 F_1 \left( l+1 + \imath \eta , 2l+2 ; - 2\imath k_e r \right)
\text{,}
\end{aligned}
\label{eq:RegCoulFun}
\end{equation}
where $\Gamma$ is the Gamma function and
$\mathstrut_1 F_1$ is the
Kummer confluent hypergeometric function.\cite{buchholz1969}
Should one be interested in studying the photodetachment of anions, the remaining core is neutral and  the asymptotic charge of $U^{\text{mol}}(r)$ is $z=0$. In this case, the Coulomb wave should be replaced by a plane wave, and the radial function  (\ref{eq:RegCoulFun}) is then a spherical Bessel function.
In reality, the main motivation for using a Coulomb wavefunction is methodological: within such an approximation, the transition matrix integrals are fully analytical, allowing for a rapid check of the convergence.

In subsequent calculations, the photoelectron is assumed to feel a distorted radial potential $U^{\text{mol}}(r)$, obtained as the angular average of the anisotropic potential calculated from the direct term in the static exchange approximation.\cite{fernandez2010,fernandez2014,granados_phd,granados2016}
At short distances, this potential behaves as $-Z_{\text{center}}/r$, where $Z_{\text{center}}$ is the approximate charge number at the central nucleus,
and features a $z=1$ Coulomb asymptotic behavior at large distances (again, in the case of photodetachment of anions, the asymptotic charge would be $z=0$). This model potential, which takes somehow into account the charge distribution of the core,  can clearly be improved as to include exchange and many-electron effects. As we shall see in section \ref{sec_results}, the potential $U^{\text{mol}}(r)$ not only provides a reasonable model that allows to capture the correct physics, it serves here also to illustrate that our approach can be applied to any central potential.

The fact that we use a central potential $U^{\text{mol}}(r)$  (pure Coulomb or distorted) thus putting aside  the multicentric nature of the target, is clearly a strong approximation.
It would not make much sense if we were interested for example in photoionization for a fixed molecular orientation.
Since in the present study we are looking at AH$_n$ molecules with a heavy central atom, and the molecules are randomly oriented as in the experiments, an angular average of the multicentric potential is not an heresy.

\subsection{\sffamily \large Complex Gaussian representation of the radial continuum wavefunction \label{subs_CGrepresentation}}

In this subsection we give an overview of the cGTO representation method introduced in ref. \citenum{ammar2020}, as a generalization of previous works on rGTOs.\cite{nestmann1990,faure2002} We also provide a table of optimal exponents used later in  section~\ref{sec_results}.

\subsubsection{\sffamily \normalsize  Fitting approach \label{subsubs_fitting}}

We consider a set of arbitrary functions $f_{p}(r)$, $p=1,\dots,p_{\text{max}}$ that we would like to approximate by linear combinations of $N$ cGTOs:
\begin{equation}
f_{p}(r) \approx f_{p}^G(r) = \sum_{i=1}^N [c_i]_{p} \exp(-\alpha_i r^2) ,
\label{eq:gaussianexpansion}
\end{equation}
where the exponents are complex, $\alpha_i = \text{Re}(\alpha_i) + i \text{Im}(\alpha_i) $, with $\text{Re}(\alpha_i)>0$.
A minimization is performed for the objective function $\Xi$ defined as 	
\begin{equation}
\begin{aligned}
&\Xi \left(\text{Re}(\alpha_1),\dots,\text{Re}(\alpha_N),
\text{Im}(\alpha_1),\dots,\text{Im}(\alpha_N)\right) \\
&=
\sum_{p} \frac{\sum_{\kappa} | f_{p}(r_{\kappa}) - f_{p}^G(r_{\kappa}) |^2}
{\sum_{\kappa} | f_{p}(r_{\kappa}) |^2}
+ D(\text{Re}(\alpha_1),\dots,\text{Re}(\alpha_N)) \text{,}
\end{aligned}
\label{FlRNC}
\end{equation}
over some given radial grid $\{r_{\kappa}\}_{{\kappa}=1,\dots,{\kappa}_{max}}$.
The $\Xi$ function depends on $2N$ non-linear real parameters, $\{\text{Re}(\alpha_i),\text{Im}(\alpha_i) \}_{i=1,\dots,N}$
(the exponents),
and $N \times p_{\text{max}}$ linear parameters
$\{[c_i]_{p}\}_{i=1,\dots,N,p=1,\dots,
		p_{max}}$
(the expansion coefficients).
In eq.~\eqref{FlRNC}, $D$ is a penalty function whose aim is to avoid  the convergence of two exponents to the same value. It is defined as
	\begin{equation}
	D(\alpha_1,\dots,\alpha_N) = \sum_{i=2}^{N} \sum_{j=1}^{i-1}
	\exp \left( -g \left| \frac{\alpha_i}{\alpha_j}
	- \frac{\alpha_j}{\alpha_i} \right| \right) \text{,}
	\label{Daddedtofit0}
	\end{equation}
where $g$ is a fixed parameter (in general $g \approx r_{\kappa_{max}}$).

We start the optimization with some reasonable set of exponents $\{\alpha_i\}$.
Based on our numerical experience, the real parts are picked between two research bounds $\text{Re}(\alpha_1) = a$ and $\text{Re}(\alpha_N) = b$ and follow the distribution
\begin{equation}
\frac{\text{Re}(\alpha_{i+1})}{\text{Re}(\alpha_{i})} = \left(\frac{b}{a} \right)^{\frac{1}{N-1}}
\text{.}
\end{equation}
This choice ensures that most exponents start with a small real part.
The imaginary parts are initially set to zero for all $\alpha_i$.

The fitting error $\Xi$ is then minimized following a two step iterative algorithm:
(i) a least square optimization gives an approximation for the coefficients $\{c_i\}$
and
(ii)
the exponents $\{\alpha_i\}$ are optimized by using the Bound Optimization BY Quadratic Approximation (BOBYQA).~\cite{powell2009}
We iterate over steps (i) and (ii) until some reasonable convergence is reached.
The BOBYQA algorithm requires research bounds that allow us to constrain the domain for the optimal exponents.

\subsubsection{\sffamily \normalsize  Optimal exponents and accuracy \label{subsubs_optexponents}}

In preparation for molecular applications, basis sets are required for several partial waves,  $l=0,1,2,3,4$ being sufficient for our purpose.
An optimization has been performed with a set of $p_{\text{max}}=6$
regular Coulomb functions with $z=1$ defined as
\begin{equation}
\mathscr{F}_{l}:\{F_{\nu}(r)=F_{l}(\eta,k_{\nu}r)\}_{\nu=1,\dots,6}
\label{eq:ensembleF}
\end{equation}
on a momentum grid $k_{\nu}=0.5+0.25(\nu-1)$, $\nu=1,\dots,6$.
Each Coulomb set $\mathscr{F}_{l}$ is fitted in a radial box $r \in [0;25]$ using $N=30$ cGTOs, \emph{i.e.}, for each value of $l$ there are $60$ non-linear real parameters to be optimized along with $30 \times 6$ linear coefficients.

The initial value for the real parts of the exponents are selected between $a = 10^{-4}$ and $ b = 10^2 $ and research bounds are chosen as to constrain $ \text{Re}(\alpha_i) \in \left[ 10^{-4},10^3\right] $ and $\text{Im}(\alpha_i) \in \left[ -0.1,0.1 \right]$.
Table \ref{tab:exponents} reports the optimal exponents for $l=0,1,2,3,4$ ordered according to their real parts.
The number of digits is necessary as to ensure accuracy.
%
%
%
%
%
%
%
\begin{figure}[htp]
\begin{center}
\includegraphics[width=0.8\linewidth]{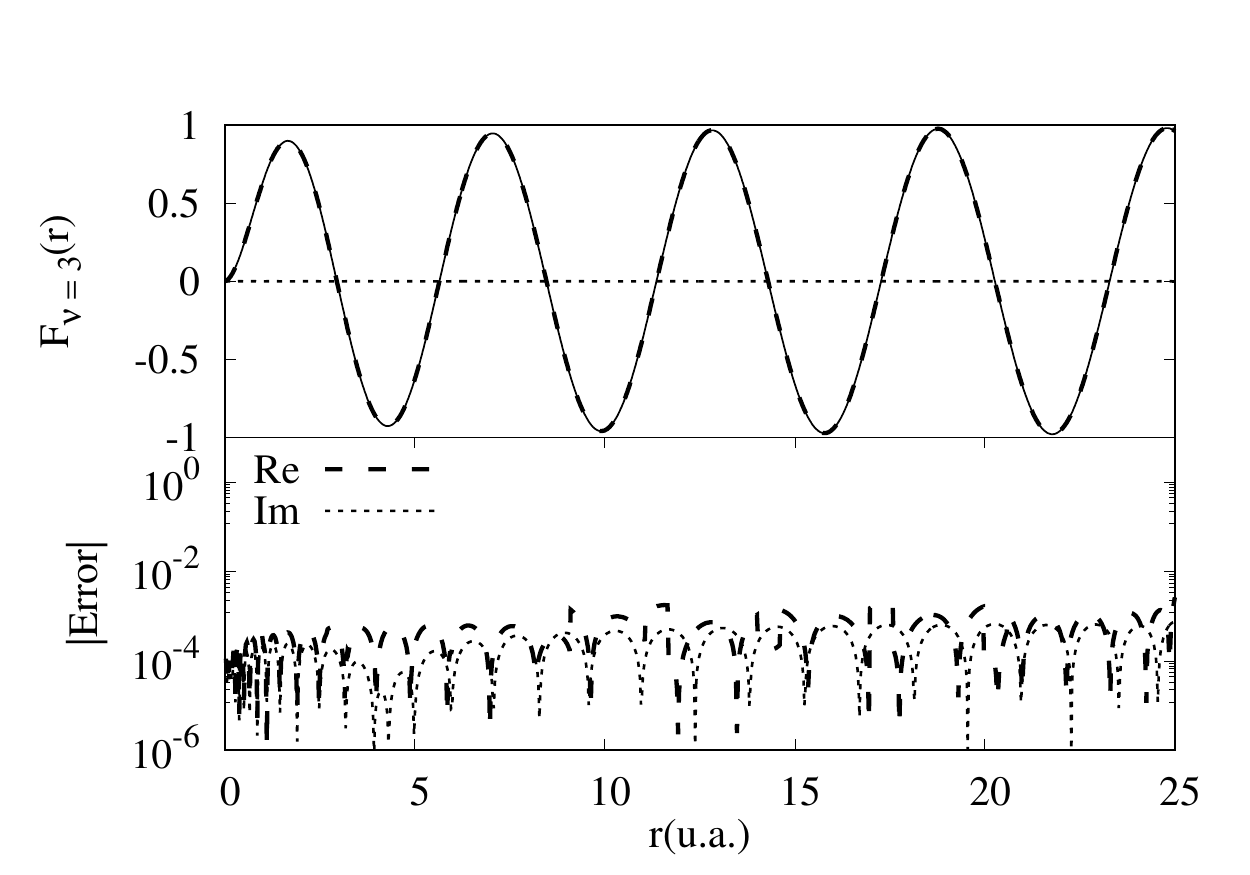}
\end{center}
\caption{Top panel : regular Coulomb function with wavenumber
 $k_e=1$ a.u.
(thin continuous line), superimposed with its cGTO fit (Real part: dashed line; null imaginary part: dotted line). The bottom panel shows the corresponding absolute errors.
}
\label{fig:fig1_fitting}
\end{figure}
The fitting error for one representative Coulomb wave
with $k_e=1$
 is shown in Figure~\ref{fig:fig1_fitting} within the radial fitting box.
Real and imaginary parts are equally well reproduced with errors less than $10^{-3}$.  As discussed in details in ref. \citenum{ammar2020}, one should also be attentive to the fitting quality outside the fitting box since some trouble may arise in case of divergences at large radial distances. We have checked that in our case the fitted functions go reasonably fast to zero at large values of $r$, thus not risking to jeopardize their application in cross-section calculations.

An optimization reaching $(\Xi - D) < 10^{-4}$ for one value of $l$ typically requires  $4-5$ hours  using 8 CPU at 2.66 GHz.
Optimizations have been made only for Coulomb functions (and not distorted waves).
However we expect that optimal exponent sets can also be used for distorted radial functions in the same energy range without significant loss of accuracy.
This will be later confirmed in the result section~\ref{sec_results}.

\subsection{\sffamily \large Closed-form expression of the cross-section using Gaussian-Slater integrals \label{subs_integrals}}

\subsubsection{\sffamily \normalsize  General expressions \label{subsubs_generalexpressions}}

We now turn to the explicit calculation of the photoionization differential cross-section and asymmetry parameter.
Let us start with a few considerations about the molecular orientation. The incident photon polarization is defined in the laboratory frame, and so is the photoelectron wavevector.
Spherical harmonics transform from the molecular frame
(from now on, unprimed variables) to the laboratory frame (from now on, primed variables) using Wigner rotation matrices,\cite{edmonds1974}
\begin{equation}
Y_l^m \left( \widehat{r'} \right)
= \sum_{\tilde m=-l}^{l}
\mathscr{D}_{ \tilde m m }^{l} \left( \widehat{\mathscr{R}} \right)
Y_l^{\tilde m} \left( \hat{r} \right).
\end{equation}
For an arbitrary molecular orientation, the dipole operator in \eqref{eq:Tlength} can be written as (for convenience we replace the modulus $r'=r$ hereafter)
\be
\begin{aligned}
\mathcal{T}^{\left( \mathscr{L} \right) }
&\equiv \hat{\epsilon} \cdot \mathbf{r'}
= r \cos{(\theta')}  \\
& = 2 \sqrt{\frac{\pi}{3}} r
Y_{1}^{0}(\widehat{r'})\\
&= \sum_{\mu=-1}^{1}
\mathscr{D}_{\mu 0}^{1}\left( \widehat{\mathscr{R}} \right)
\mathcal{T}_{\mu}^{\left( \mathscr{L} \right) }, \\
\end{aligned}
\label{eq:rotated_op1}
\ee
with
\be
\mathcal{T}_{\mu}^{\left( \mathscr{L} \right) }
= 2 \sqrt{\frac{\pi}{3}} r  Y_{1}^{\mu}(\hat{r}).
\label{eq:TmuL}
\ee
A similar rotated expression holds for the gradient operator in velocity form \eqref{eq:Tvelocity},
\begin{equation}
\begin{aligned}
\mathcal{T}^{\left( \mathscr{V} \right) }
& \equiv
\frac{1}{\imath E_{\gamma}}\hat{\epsilon} \cdot \mathbf{p}  \\
& = \sum_{\mu=-1}^{1}
\mathscr{D}_{\mu 0}^{1}\left( \widehat{\mathscr{R}} \right)
\mathcal{T}_{\mu}^{\left( \mathscr{V} \right) },
\end{aligned}
\label{eq:rotated_op2}
\end{equation}
with
\be
\mathcal{T}_{\mu}^{(\mathscr{V})}
=
-\frac{1}{E_{\gamma}}  \nabla_{\mu},
\label{eq:TmuV}
\ee
$\nabla_{\mu}$ being the spherical tensor components of the gradient operator (see section 5 of ref. \citenum{edmonds1974}).
We should also take into account the inverse rotation matrix between
the molecular and the laboratory frame since the photoelectron
wavevector  is measured in the laboratory frame while  $\phi_i$ in eq. \eqref{eq:Moccia_phi} corresponds to a given molecular orientation. Thus the spherical harmonic in eq. \eqref{eq:psi_f} is written as
$Y_l^{m*}(\widehat{k_e}) = \sum_{\tilde m=-l}^{l}
\mathscr{D}_{ m \tilde m }^{l} \left( \widehat{\mathscr{R}} \right)
Y_l^{\tilde m *} \left( \widehat{k_e'} \right)$ with $\widehat{k_e'}$ the wavevector direction in the laboratory frame.

Following Chandra\cite{chandra1987} and in the case of a linearly polarized photon, the explicit formula of \eqref{eq:PICSdiff} after averaging over all possible Euler angles becomes
\be
\frac{d \sigma^{\left(\mathscr{G}\right)} \left(\widehat{k_e'} \right) }{d \widehat{k_e'}}
= \sum_{L} \mathcal{A}_L^{\left( \mathscr{G} \right)} \left( k_e \right)
 \, Y_L^{0} \left( \widehat{k_e'} \right)
\label{eq:dsigmachandra}
\ee
where
\begin{equation}
\begin{aligned}
\mathcal{A}_L^{\left( \mathscr{G} \right)} \left( k_e \right)
& \equiv
\frac{8 \pi E_{\gamma}}{k_e \, c}
\sum_{l_1,m_1,\mu_1} \,
\sum_{l_2,m_2,\mu_2} \,
(-\imath)^{l_1-l_2}  e^{\imath \left( \delta_{l_1} - \delta_{l_2} \right) }
\mathcal{M}_{l_1,m_1,\mu_1}^{\left( \mathscr{G} \right)}
\left( \mathcal{M}_{l_2,m_2,\mu_2}^{\left( \mathscr{G} \right)} \right)^* \\
& \quad \times
(-1)^{m_1-\mu_1}  \sqrt{\frac{(2l_1+1)(2l_2+1)(2L+1)}{4 \pi}}
\tj{l_1}{l_2}{L}{0}{0}{0} \tj{1}{1}{L}{0}{0}{0}  \\
& \quad \times
\tj{1}{1}{L}{\mu_1}{-\mu_2}{-\mu_1+\mu_2}
\tj{l_1}{l_2}{L}{-m_1}{m_2}{\mu_1-\mu_2} ,
\end{aligned}
\label{eq:ALke}
\end{equation}
and $\mathcal{M}_{l_1,m_1,\mu_1}^{\left( \mathscr{G} \right)}$ stand for the integrals
\be
\mathcal{M}_{l,m,\mu}^{\left( \mathscr{G} \right)} \equiv
\int  \left( \frac{u_{l,k_e}(r)}{r} Y_l^m(\hat{r}) \right)^*
\mathcal{T}_{\mu}^{\left( \mathscr{G} \right) }
\phi_i({\bf{r}})
d{\bf{r}} .
\label{eq:integrals}
\ee
In \eqref{eq:ALke}, the superscript $\mathscr{G}$ refers to the gauge choice,
the phase shifts
 $\delta_{l_{1}}$ and $\delta_{l_{2}}$ come from the partial wave expansion  \eqref{eq:psi_f},
and the standard notation is used for Wigner $3j$ symbols.\cite{edmonds1974}

The summation in \eqref{eq:dsigmachandra} actually involves  only two non-zero terms,
 $L=0$ and $L=2$, and
is thus equivalent to expression \eqref{eq:asymmetry_parameter} written in terms of the asymmetry parameter.
Using the above notations,
the total cross-section and asymmetry parameter can be respectively calculated as
\begin{equation}
\begin{aligned}
\sigma^{\left(\mathscr{G}\right)} (k_e)
=& \int
\left[
\mathcal{A}_0^{\left( \mathscr{G} \right)} (k_e) Y_0^0 \left(\widehat{k_e'} \right)
+
\mathcal{A}_2^{\left( \mathscr{G} \right)} (k_e) Y_2^0 \left(\widehat{k_e'} \right)
\right]
 d \widehat{k_e'} \\
=& \sqrt{4 \pi} \, \mathcal{A}_0^{\left( \mathscr{G} \right)} (k_e) \\
=& \frac{8 \pi E_{\gamma}}{3 \, k_e \, c}
\sum_{l,m,\mu}
\left| \mathcal{M}_{l,m,\mu}^{\left( \mathscr{G} \right)} \right|^2,
\end{aligned}
\label{eq:restotalcrosssection}
\end{equation}
and
\begin{equation}
\begin{aligned}
\beta^{\left(\mathscr{G}\right)} (k_e) &=  \sqrt{20 \pi} \,
\frac{\mathcal{A}_2^{\left( \mathscr{G} \right)} (k_e) }
	{\sigma^{\left(\mathscr{G}\right)} (k_e) } \\
&= \sqrt{5} \, \frac{\mathcal{A}_2^{\left( \mathscr{G} \right)} (k_e)}
	{\mathcal{A}_0^{\left( \mathscr{G} \right)} (k_e)},
\end{aligned}
\label{eq:resasymmetry}
\end{equation}
with $\mathcal{A}_{2}^{\left( \mathscr{G} \right)} (k_e)$ given by \eqref{eq:ALke}.

The numerical calculation of \eqref{eq:restotalcrosssection} and \eqref{eq:resasymmetry} is greatly facilitated by the fact that integrals \eqref{eq:integrals} becomes analytical by using cGTOs expansions for the continuum radial function $u_{l,k_e}(r)$.

\subsubsection{\sffamily \normalsize  Transition integrals in length or velocity gauges}

In the length gauge, we first insert the transition operator \eqref{eq:TmuL} in integral \eqref{eq:integrals},
and then make use of the initial state expansion \eqref{eq:Moccia_phi}. The angular part yields
the Gaunt coefficients\cite{gaunt1929,edmonds1974} 
denoted by
$
\braket{l_f \, m_f | L \, M | l_i \, m_i}
\equiv \int
Y_{l_f}^{m_f *} \left( \hat{r} \right)
Y_{L}^{M} \left( \hat{r} \right)
Y_{l_i}^{m_i} \left( \hat{r} \right) \,  d \hat{r}
$.
We thus obtain
\begin{equation}
\begin{aligned}
\mathcal{M}_{l,m,\mu}^{\left( \mathscr{L} \right)}
&= 2 \sqrt{\frac{\pi}{3}} \sum_{j=1}^{N_i} C_{ij}
\braket{ l,m |1,\mu|l_j,m_j }
 \mathcal{I}_{j,l}^{(2)}
\end{aligned}
\label{eq:Melementlenght}
\end{equation}
where the radial integral
\begin{equation}
\begin{aligned}
\mathcal{I}_{j,l}^{(g)}
&=  \int_0^{\infty} \left(u_{l,k_e}(r)\right)^* e^{-\zeta_j r} r^{n_j+g-1} dr ,
\end{aligned}
\label{eq:integ_rad_g}
\end{equation}
is defined here for a given integer $g$.
Note that the sum over $j$ in \eqref{eq:Melementlenght} refers to different contributions in the initial state expansion \eqref{eq:Moccia_phi}; each value of $j$ is associated with  a pair $(n_j, \zeta_j)$.
The Gaunt coefficient imposes the restrictions $l=l_j \pm 1$ and $m=\mu+m_j$, selecting therefore some specific terms in this expansion.

The formulation is a little bit more complicated in velocity gauge.
First, the gradient operator \eqref{eq:TmuV} is inserted in \eqref{eq:integrals}, leading to
\begin{equation}
\begin{aligned}
\mathcal{M}_{l,m,\mu}^{\left( \mathscr{V} \right)}
&=
- \frac{1}{E_{\gamma}} \sum_{j=1}^{N_i} C_{ij}
\int_0^{\infty} r^2 dr \frac{\left( u_{l,k_e}(r) \right)^*}{r}
\braket{l,m \left| \nabla_{\mu} \left( r^{n_j-1} e^{-\zeta_j r} \right) \right| l_j,m_j  }.
\end{aligned}
\label{eq:Melementvelocity}
\end{equation}
Replacing the gradient formula given in section 5.7 of ref. \citenum{edmonds1974}, the angular integral in \eqref{eq:Melementvelocity} becomes
\begin{equation}
\begin{aligned}
\braket{l \, m| \nabla_{\mu} \left( r^{n_j-1} e^{-\zeta_j r} \right) |l_j \, m_j}
&= \Pi_{l \, 1 \,l_j}^{m \, \mu \,m_j}  \, \Delta_{l \, l_j}
\left(\partial_r + \frac{b_{l \, l_j}}{r}\right) \left( r^{n_j-1} e^{-\zeta_j r} \right) \\
&=
\Pi_{l \, 1 \,l_j}^{m \, \mu \,m_j}  \,
\Delta_{l \, l_j}
\left[ -\zeta_j r  + \left( n_j-1 + b_{l \, l_j} \right)  \right] r^{n_j-2}  e^{-\zeta_j r}
\end{aligned}
\end{equation}
with angular coefficients
\be
\Pi_{l \, 1 \,l_j}^{m \, \mu \,m_j} = (-1)^{m} \tj{l}{1}{l_j}{-m}{\mu}{m_j} \tj{l}{1}{l_j}{0}{0}{0}^{-1} ,
\ee
\begin{equation}
\Delta_{l \, l_j}  =
\left\lbrace
\begin{aligned}
&\frac{l_j+1}{\sqrt{(2l_j+1)(2l_j+3)}} \quad \text{if } l =  l_j+1 \text{,} \\
&\frac{l_j}{\sqrt{(2l_j-1)(2l_j+1)}} \quad \text{if } l = l_j-1 \text{,}
\end{aligned}
\right.
\end{equation}
and
\begin{equation}
b_{l \, l_j}  =
\left\lbrace
\begin{aligned}
&-l_j \quad \text{if } l =  l_j+1 \text{,} \\
&l_j+1 \quad \text{if } l = l_j-1 \text{.}
\end{aligned}
\right.
\end{equation}
Using again the notation \eqref{eq:integ_rad_g}, the dipole integral becomes
\begin{equation}
\begin{aligned}
\mathcal{M}_{l,m,\mu}^{\left( \mathscr{V} \right)}
&=
- \frac{1}{E_{\gamma}}
\sum_{j=1}^{N_i} C_{ij}
\Pi_{l \, 1 \,l_j}^{m \, \mu \,m_j}
\Delta_{l \, l_j}
\left[  -\zeta_j  \mathcal{I}_{j,l}^{(1)} +
\left( n_j-1 + b_{l \, l_j} \right) \mathcal{I}_{j,l}^{(0)}  \right]. \\
\end{aligned}
\label{eq:Melementvelocity2}
\end{equation}
Now let us assume that the continuum radial function,
as defined in \eqref{eq:fct_disto_DE} or \eqref{eq:RegCoulFun},
is expanded in an optimal cGTO set of the form \eqref{eq:gaussianexpansion},
\emph{i.e.},
\begin{equation}
\left( u_{l,k_e}(r) \right)^* \approx \sum_{s=1}^{N} \left[ c_s \right]_{l,k_e} e^{-\left[ \alpha_s \right]_{l} r^2}
\text{.}
\label{eq:u_Gauss}
\end{equation}
Then radial integrals \eqref{eq:integ_rad_g} can be written in closed form as
\begin{equation}
\begin{aligned}
\mathcal{I}_{j,l}^{(g)}
&=
\sum_{s=1}^N
 \int_0^{\infty} \left[ c_s \right]_{l,k_e} e^{-\left[ \alpha_s \right]_{l} r^2} e^{-\zeta_j r} r^{n_j+g-1} dr  \\
	= & \frac{(n_j+g-1)!}{2^{n_j+g}} \sum_{s=1}^N
	\frac{\left[ c_s \right]_{l,k_e}}
	{\left[ \alpha_s \right]_{l}^{\frac{n_j+g}{2}}}
	\, U \left( \frac{n_j+g}{2} , \frac{1}{2};
	\frac{\zeta_j^2}{4 \left[ \alpha_s \right]_{l}} \right)
	\text{,}
	\end{aligned}
\label{eq:Z_FGC}
\end{equation}
where
$U(\cdot ,\cdot ; \cdot)$ is the Tricomi function.\cite{bateman1953,gradshteyn2014}
It can be numerically evaluated, for example,
using the Python library \emph{mpmath}.\cite{mpmath}
Expression \eqref{eq:Z_FGC} depends on the optimal complex exponents $\{\alpha_s\}$ and expansion coefficients $\{c_s\}$.
 Note that if the radial continuum function $u_{l,k_e}(r)$ is real, the complex conjugation in \eqref{eq:integ_rad_g} has no effect.
However, in the case its cGTOs expansion is not perfectly real, we approximate the integral to its real part, $\mathcal{I}_{j,l}^{(g)} \simeq \text{Re}(\mathcal{I}_{j,l}^{(g)})$.

In the particular case of a pure Coulomb continuum function
given by \eqref{eq:RegCoulFun} with $\eta=-\frac{1}{k_e}$, it is even possible to compute integrals \eqref{eq:integ_rad_g} in closed form without using cGTO expansions of the radial function.
Indeed, we have
\begin{equation}
\begin{aligned}
\mathcal{I}_{j,l}^{(g)\text{Coul}}
&=
\int_0^{\infty}
F_l\left(\eta, k_e r \right )
e^{-\zeta_j r} r^{n_j+g-1} dr \\
 &=
k_e^{l+1}
e^{\frac{\pi}{2k_e}}
\frac{2^{l} \left| \Gamma\left(l + 1 -\frac{\imath}{k_e} \right) \right|}
{\Gamma \left( 2 l  + 2 \right)}
\frac{\Gamma \left( n_j+g+l +1\right) }{(\zeta_j - \imath k_e)^{n_j+g+l+1}} \\
 \times
 \mathstrut_2 F_1 & \left(  l + 1 +\imath\eta
, n_j+g+l +1 , 2 l  + 2 ; \frac{-2 \imath k_e}{\zeta_j - \imath k_e} \right)
\end{aligned}
\label{eq:Z_coulomb}
\end{equation}
where $\, \mathstrut_2 F_1$ is the Gauss hypergeometric function  (see formula 13.10.3 of ref. \citenum{NISTdigital}).
This analytical result can be easily evaluated.\cite{mpmath} It can be used as a benchmark to validate the evaluation of \eqref{eq:Z_FGC} that makes use of the cGTO representation.

Before presenting our results, we should emphasize that in ref. \citenum{ammar_phd} the photoionization of atomic hydrogen was considered and used as a testbed for the analytical evaluation of matrix elements. Indeed, the cross-section from the exact initial states $n_i l_i m_i$ is known in closed form, and formulae \eqref{eq:Z_FGC} and \eqref{eq:Z_coulomb} could be tested successfully.

\section{\sffamily \Large RESULTS \label{sec_results}}

We shall now present partial cross-sections $\sigma(k_e)$ and asymmetry parameters $\beta(k_e)$ for NH$_3$ and H$_2$O photoionization.
The initial state is described with Moccia's monocentric expansion as explained in subsection \ref{subs_wavefunctions} and the final continuum state is represented with cGTO optimal sets as introduced in subsection \ref{subs_CGrepresentation}.
In each case, assuming that the cGTO set has been previously optimized, the numerical evaluation of closed-form expressions given by eqs. \eqref{eq:restotalcrosssection},
\eqref{eq:resasymmetry},
\eqref{eq:Melementlenght} and \eqref{eq:Melementvelocity2}
for one value of $\sigma(k_e)$  and $\beta(k_e)$ in the two gauges takes about 2 minutes of computational time using one CPU at 2.67 GHz.
For each molecule, we first validate the cGTO approach using internal comparisons for one selected valence orbital, {\emph{i.e.}}, we compare the results obtained using cGTO representations for the radial continuum function with results given by the original wavefunction.
We do this for the outgoing electron described by either a pure Coulomb wave or a distorted wave, and in both velocity and length gauges.
Then, we present the distorted wave results for cross-sections and asymmetry parameters for some other orbitals, along with comparisons with experimental and other theoretical results from the literature.

\subsection{\sffamily \large NH$_3$ photoionization parameters \label{subs_NH3}}

\begin{figure}[htp]
\begin{center}
\includegraphics[width=0.7\linewidth]{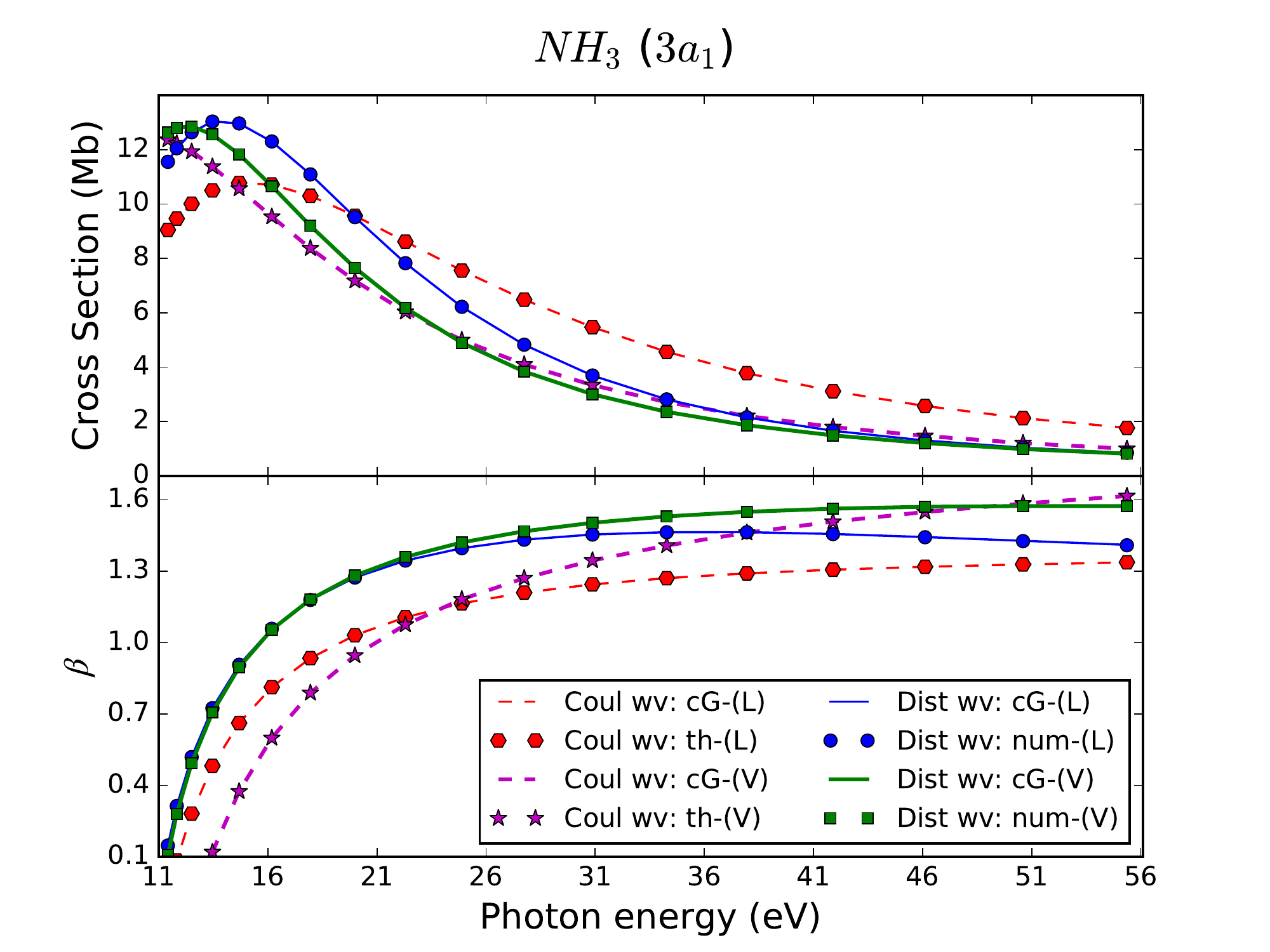}
\end{center}
\caption{Partial cross-section $\sigma(k_e)$ (top panel) and asymmetry parameter $\beta(k_e)$ (bottom panel) as a function of the photon energy (in eV), for the outer valence orbital $3a_1$ of NH$_3$.
Results using cGTOs integrals (cG) are compared with results from exact integrals (th) in the Coulomb wave case and with results from numerical integrals (num) in the distorted wave case. Calculations are performed in both length (L) and velocity (V) gauges.
}
\label{fig:NH3internal}
\end{figure}

The ground state electronic configuration of NH$_3$ is $1a_1^2 \; 2a_1^2 \; 1e^4 \; 3a_1^2 \; \; ^1A_1 $.
Fig. \ref{fig:NH3internal} shows the partial cross-section and asymmetry parameter for the lowest energy ionization ($3a_1$ valence orbital), obtained using either optimal cGTO representation for the radial continuum wavefunction, or the original wavefunction.
Results are shown for length and velocity gauges, and using either a pure Coulomb wave or a distorted wave as explained in subsection \ref{subs_wavefunctions}.
In the Coulomb wave case, we compare the results obtained using cGTOs and Gaussian integrals (eq. \eqref{eq:Z_FGC}) with those obtained by direct Coulomb integrals (eq. \eqref{eq:Z_coulomb}). The curves being almost undistinguishable, the quality of the cGTO representation is clearly sufficient to reproduce the exact results, in both length and velocity gauges.
In the case of a distorted radial function, the comparison of cGTO results with those obtained by numerical integration of the original distorted continuum wave, leads
to the same conclusion.
The results from the two continuum models, Coulomb wave {\textit{vs}} distorted wave, are expectedly different,
illustrating the strong sensitivity of the results to the physical model used to describe the outgoing electron wavefunction. 
Besides, the observed gauge dependence should not surprise since both initial and final states are not exact eigenfunctions of the Hamiltonian.
 The important point here is that, from a methodological point of view,  the Gaussian approach is validated in both cases, and in both gauges. In other words, using cGTOs does not jeopardize the quality of photoionization parameters within the limits of the selected model.

\begin{figure}[htp]
\begin{center}
\includegraphics[width=0.7\linewidth]{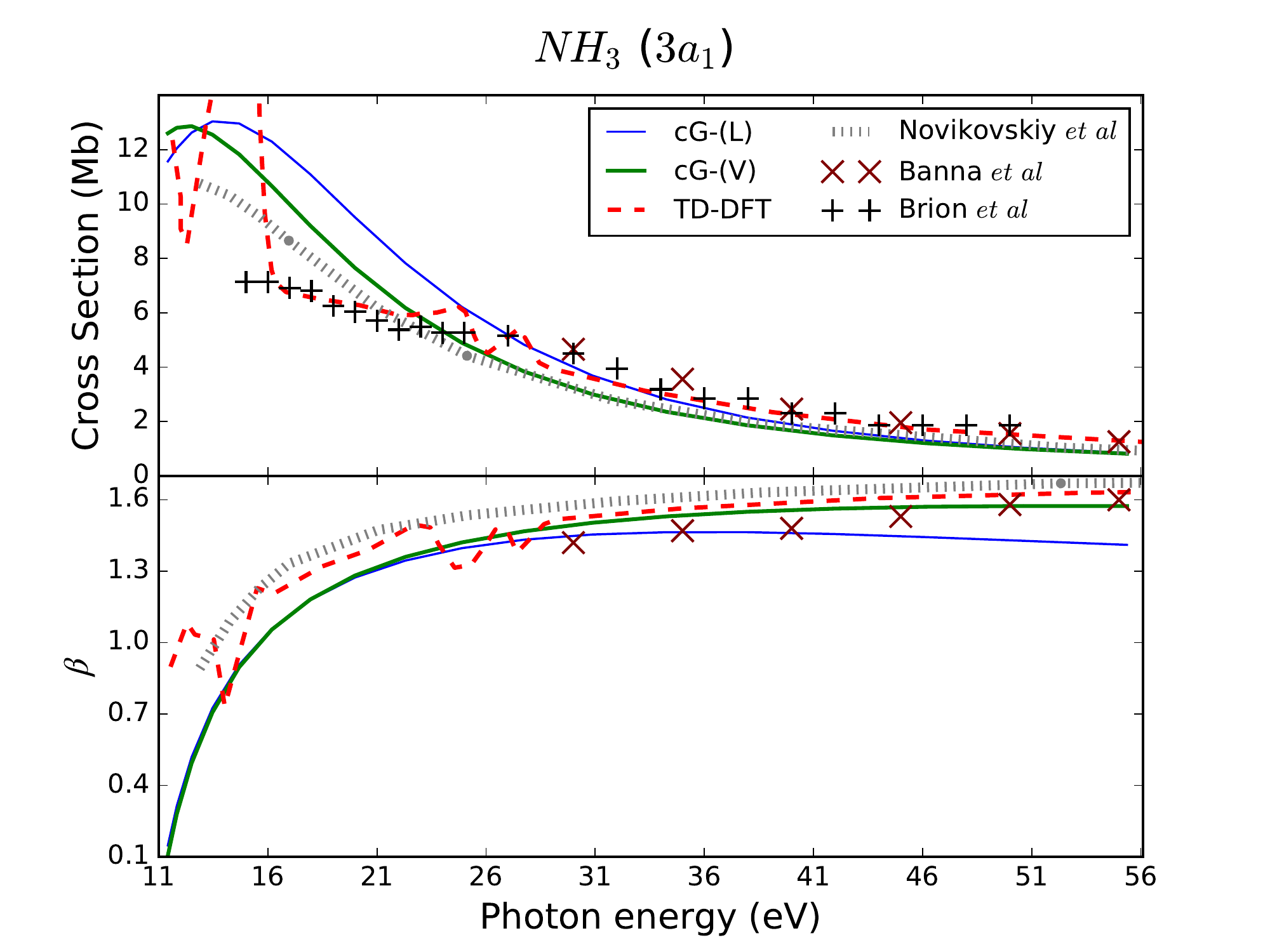}
\end{center}
\caption{Partial cross-section $\sigma(k_e)$ (top panel) and asymmetry parameter $\beta(k_e)$ (bottom panel) as a function of the photon energy (in eV), for orbital $3a_1$
of NH$_3$.
Present results using cGTOs integrals (cG) with a distorted continuum wave are compared with results from other theoretical methods
(TD-DFT by Stener {\textit{et al}}\cite{stener2002} and single-center method of Novikovskiy {\textit{et al}}\cite{novikovskiy2019}),
and experimental points
(Brion {\textit{et al}}\cite{brion1977} and Banna {\textit{et al}}\cite{banna1987}).
Our calculations are performed in both length (L) and velocity (V) gauges.
}
\label{fig:NH3comp3a1}
\end{figure}

\begin{figure}[htp]
\begin{center}
\includegraphics[width=0.7\linewidth]{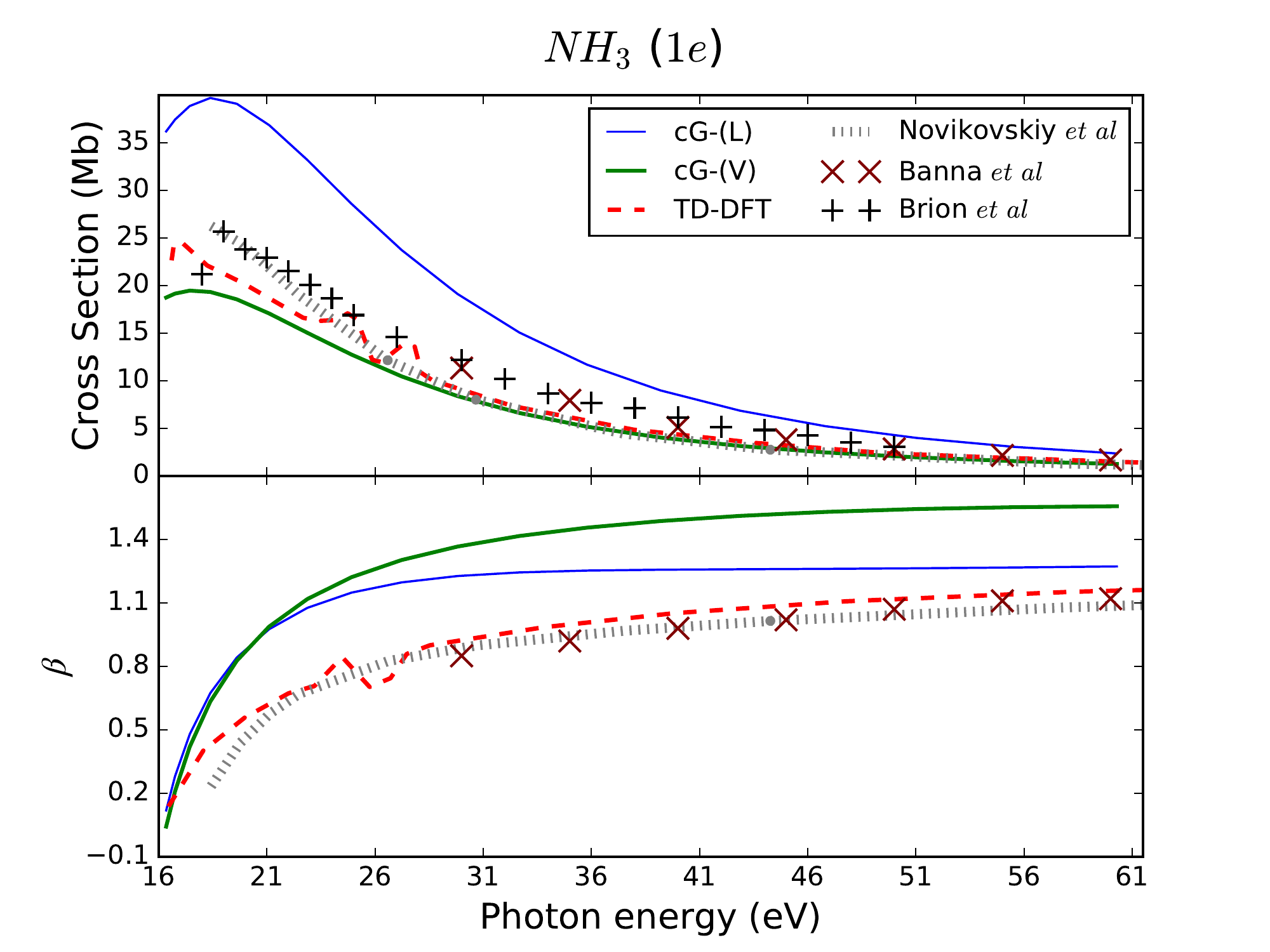}
\end{center}
\caption{Same as Fig. \ref{fig:NH3comp3a1} for  orbital $1e$ of NH$_3$.
}
\label{fig:NH3comp1e}
\end{figure}

\begin{figure}[htp]
\begin{center}
\includegraphics[width=0.7\linewidth]{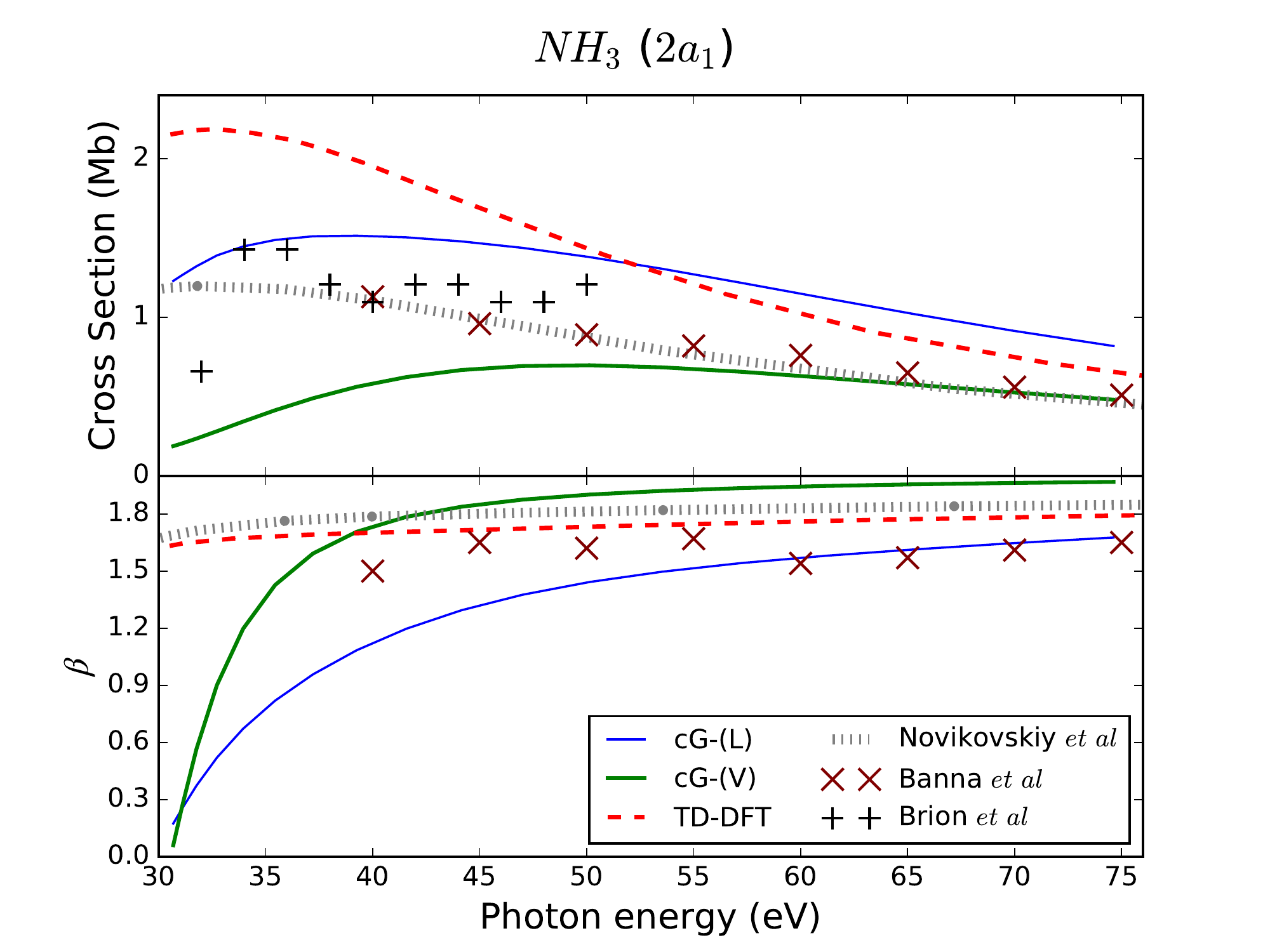}
\end{center}
\caption{Same as Fig. \ref{fig:NH3comp3a1} for  orbital $2a_1$ of NH$_3$.
}
\label{fig:NH3comp2a1}
\end{figure}

To put our results in perspective with respect to the literature, Fig. \ref{fig:NH3comp3a1}, \ref{fig:NH3comp1e} and \ref{fig:NH3comp2a1} show the partial cross-sections and asymmetry parameters for ionization from orbitals
having the lowest ionization energies, {\emph{i.e.}} from orbitals $3a_1$, $1e$ and $2a_1$, respectively.
Results are compared with other theoretical results, some of them using more sophisticated models,\cite{stener2002,novikovskiy2019} and with experimental points.\cite{brion1977,banna1987}
Note that experimental results are typically associated with uncertainties of about $3-5\%$ (not shown in the figures).
We get an overall reasonable agreement for the partial cross-sections, given the simplicity of the monocentric model.
The difference between length and velocity gauge results is sometimes quite large but the experimental cross-sections are always framed between the two results. A meaningful comparison with other theoretical cross-sections should be done within the same gauge, \emph{i.e.} in velocity gauge for ref. \citenum{novikovskiy2019},
while in length gauge for ref. \citenum{stener2002}.
Results for the asymmetry parameter, which is related to the photoelectron  angular distribution and thus is a more sensitive quantity, are good for the outer valence orbital $3a_1$ but larger deviations are visible in the $1e$ case.  For the $2a_1$ orbital, the model yields less satisfactory results.
We thus observe two general trends: (i) our results for outer orbitals give an overall better agreement with experiments and other calculations, for both the cross section and the asymmetry parameter and (ii) cross section results are in general better than the asymmetry results.
We think that these variations are mainly due to the simplicity of the static-exchange potential (used in the one-active-electron model for the description of the final state) which is  not sufficiently realistic for more internal target orbitals.
Moccia's wavefunctions describing the initial bound orbitals could also be suspected to be responsible for such variations. However, this can be ruled out by the fact that the results of ref. \citenum{novikovskiy2019}, using a similar level of approximation (a single-center Hartree-Fock wavefunction) for the initial state
but a different model for the final state, differ from ours.
Once again, the aim of the present study is not to defeat  previous results from the literature in terms of physical description but rather to prove the feasibility of fast cGTO calculations in the context of molecular photoionization processes. In this respect, the present results are fully  satisfactory.

\subsection{\sffamily \large H$_2$O photoionization parameters \label{subs_H2O}}

\begin{figure}[htp]
\begin{center}
\includegraphics[width=0.7\linewidth]{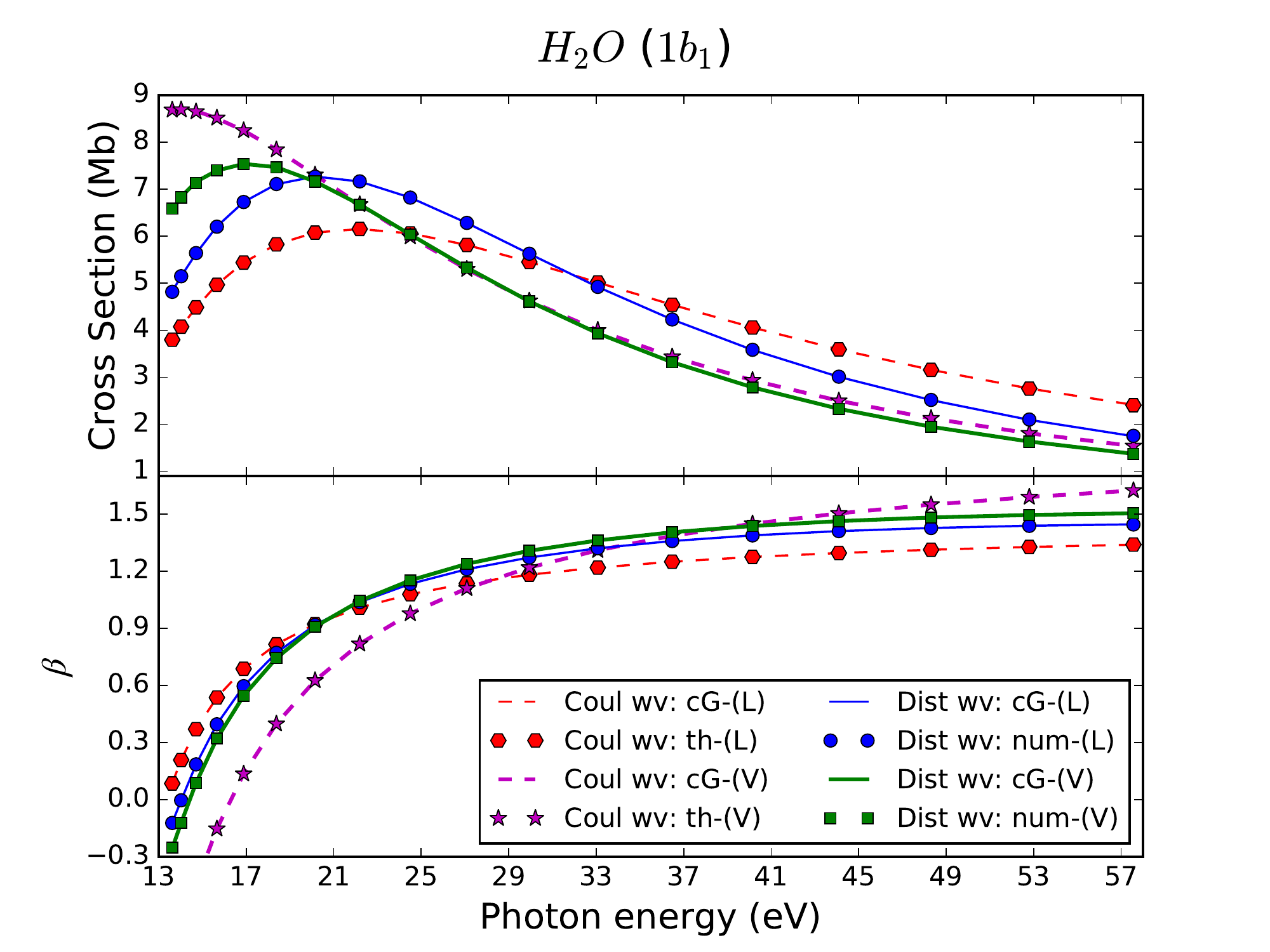}
\end{center}
\caption{Partial cross-section $\sigma(k_e)$ (top panel) and asymmetry parameter $\beta(k_e)$ (bottom panel) as a function of the photon energy (in eV), for the outer valence orbital $1b_1$ of H$_2$O.
Results using cGTOs integrals (cG) are compared with results from exact integrals (th) in the Coulomb wave case and with results from numerical integrals (num) in the distorted wave case. Calculations are performed in both length (L) and velocity (V) gauges.
}
\label{fig:H2Ointernal}
\end{figure}

We now turn to the results for water.
The ground state electronic structure of H$_2$O is
$1a_1^2 \; 2a_1^2 \; 1b_2^2 \; 3a_1^2 \; 1b_1^2 \; \; ^1A_1$.
An internal comparison between the partial cross-section and asymmetry parameter (for the outer valence orbital $1b_1$) obtained using the original functions or their cGTO representation,
is shown in Fig. \ref{fig:H2Ointernal}.
As previously observed for ammonia, the agreement between cGTO calculations and results from either theoretical integrals (in the Coulomb wave approximation) or numerical integrals (in the distorted wave case) is
 very good. And this holds true in both length and velocity gauges.

\begin{figure}[htp]
\begin{center}
\includegraphics[width=0.7\linewidth]{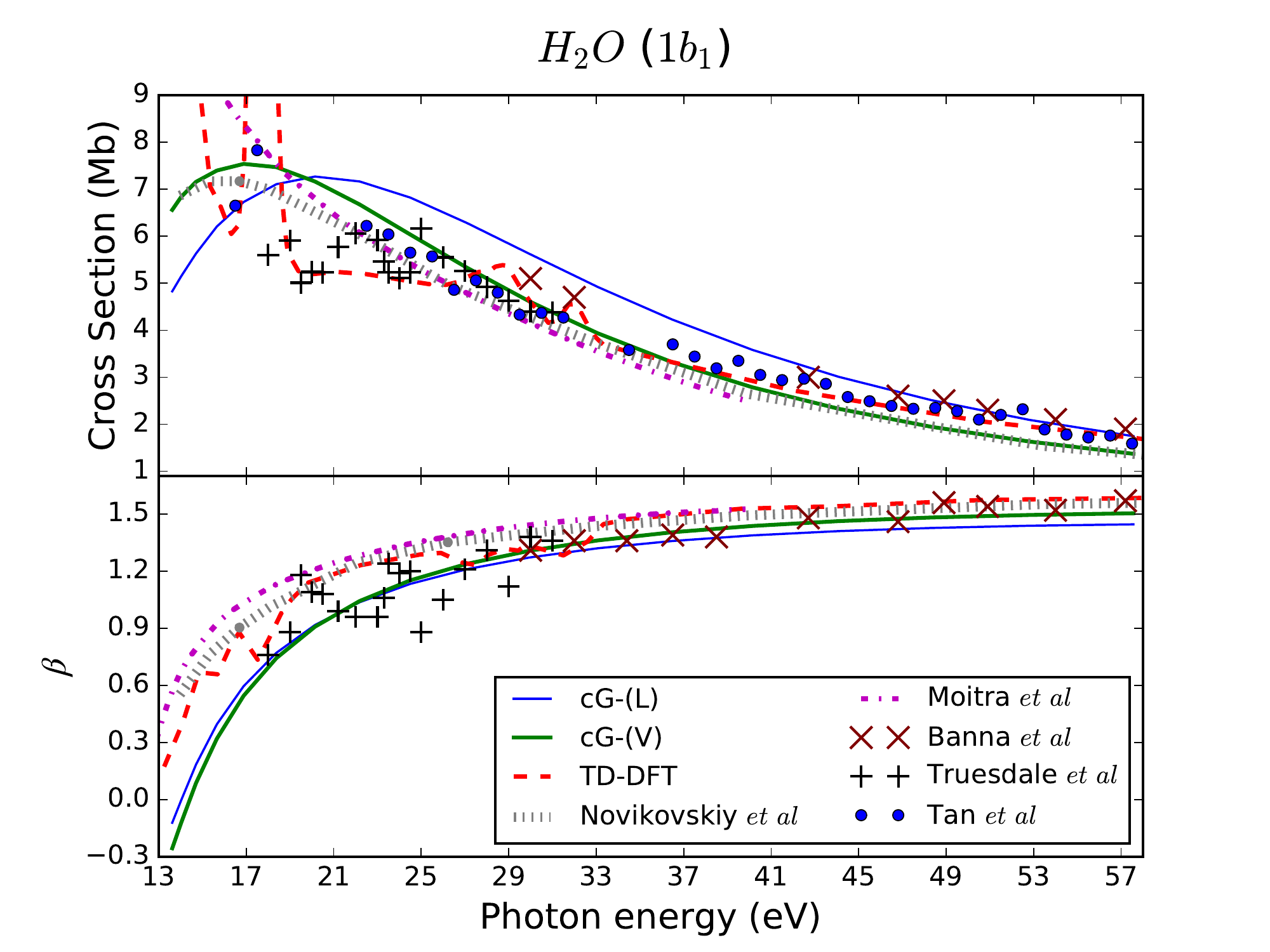}
\end{center}
\caption{Partial cross-section $\sigma(k_e)$ (top panel) and asymmetry parameter $\beta(k_e)$ (bottom panel) as a function of the photon energy (in eV), for orbital $1b_1$
of H$_2$O.
Present results using cGTOs integrals (cG) with a distorted continuum wave are compared with results from other theoretical methods
(TD-DFT by Stener {\textit{et al}}\cite{stener2002}, single-center method of Novikovskiy {\textit{et al}}\cite{novikovskiy2019}
and
EOM-CCSD Dyson/B-spline DFT calculations of Moitra {\textit{et al}}\cite{moitra2020}),
and experimental points
(Banna {\textit{et al}}\cite{banna1986},
Truesdale {\textit{et al}}\cite{truesdale1982} and
Tan {\textit{et al}}\cite{tan1978}).
Our calculations are performed in both length (L) and velocity (V) gauges. }
\label{fig:H2Ocomp1b1}
\end{figure}

\begin{figure}[htp]
\begin{center}
\includegraphics[width=0.7\linewidth]{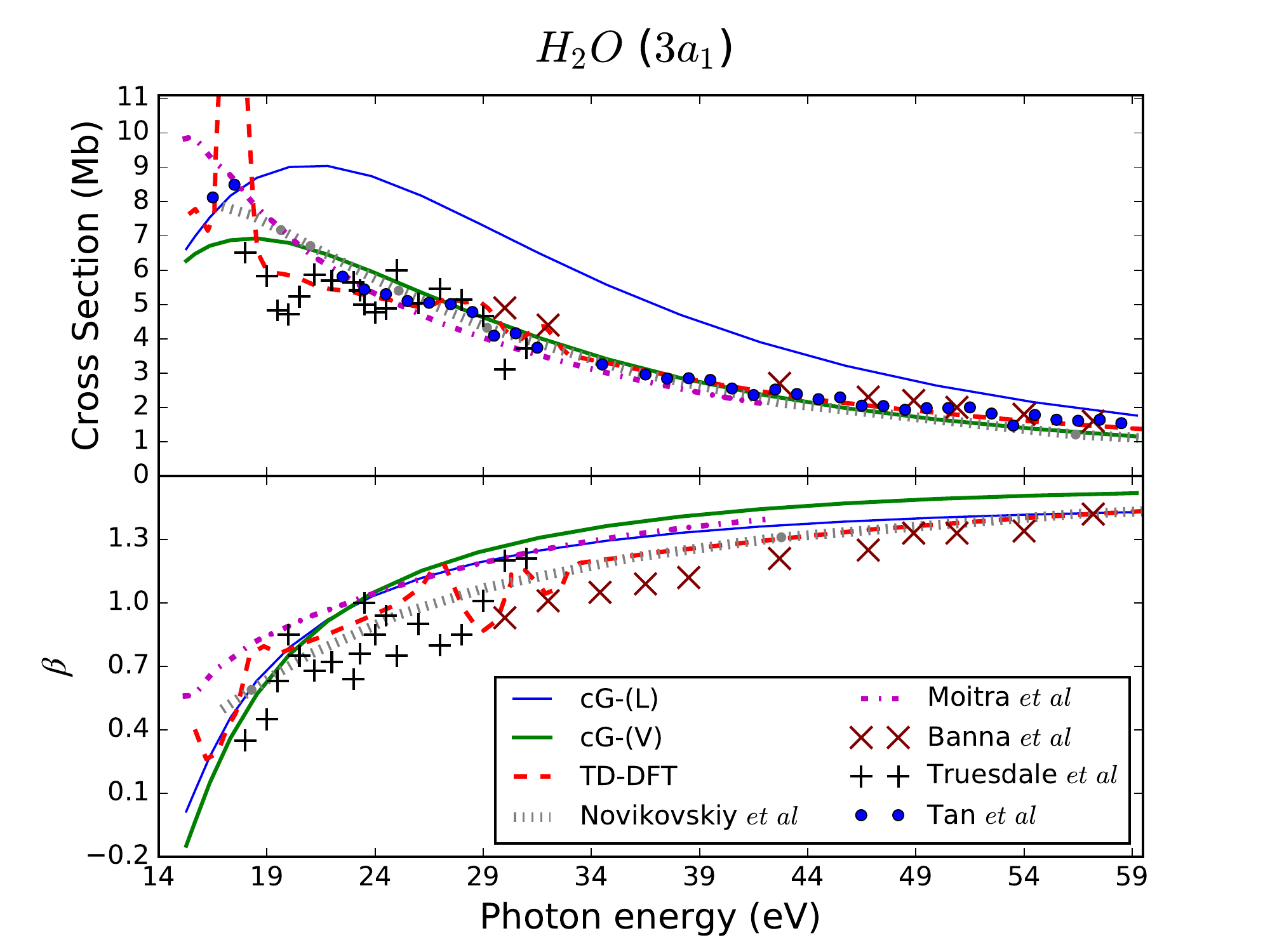}
\end{center}
\caption{Same as Fig. \ref{fig:H2Ocomp1b1} for  orbital $3a_1$ of H$_2$O.
}
\label{fig:H2Ocomp3a1}
\end{figure}

\begin{figure}[htp]
\begin{center}
\includegraphics[width=0.7\linewidth]{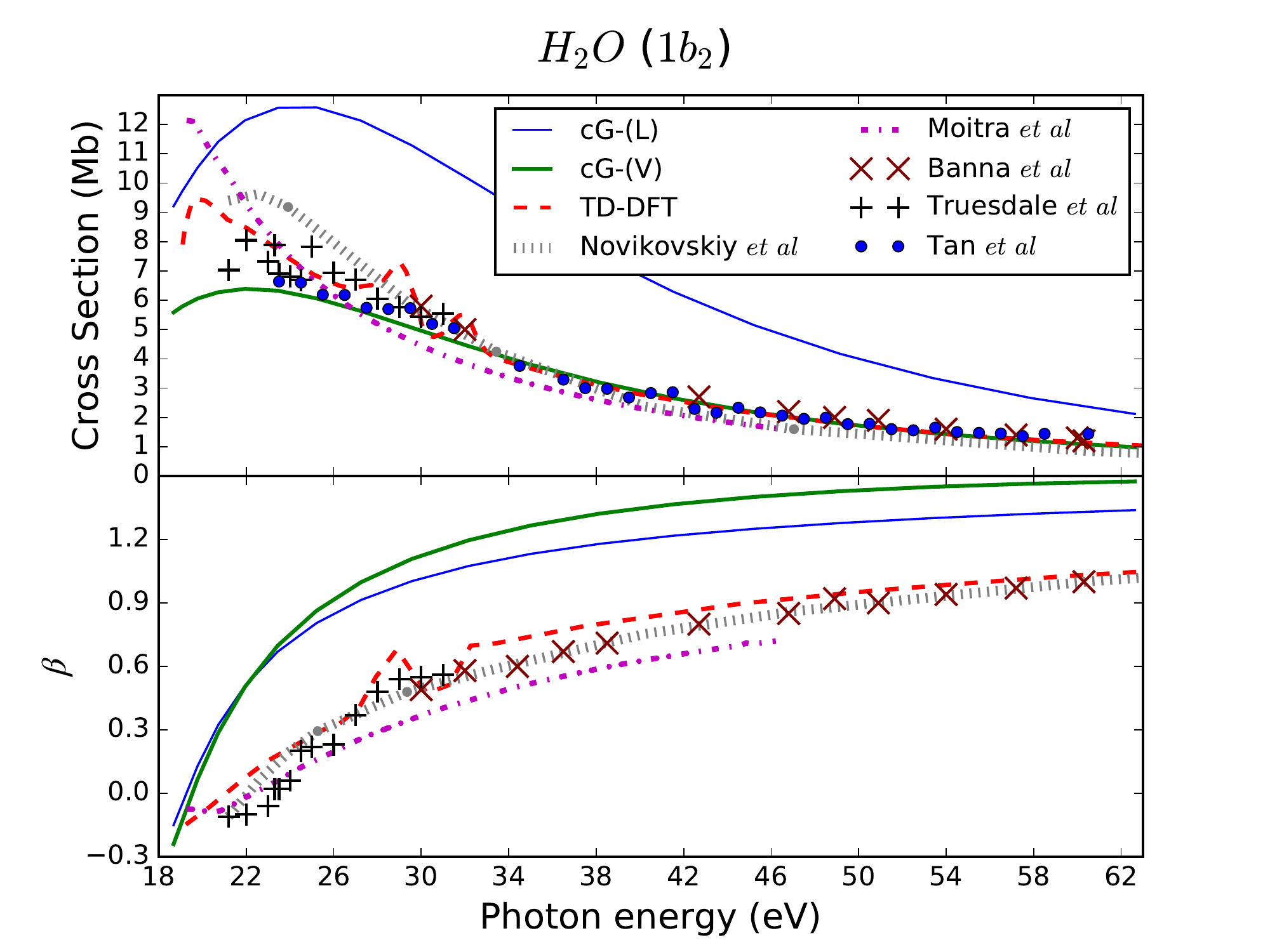}
\end{center}
\caption{Same as Fig. \ref{fig:H2Ocomp1b1} for  orbital $1b_2$ of H$_2$O.
}
\label{fig:H2Ocomp1b2}
\end{figure}

\begin{figure}[htp]
\begin{center}
\includegraphics[width=0.7\linewidth]{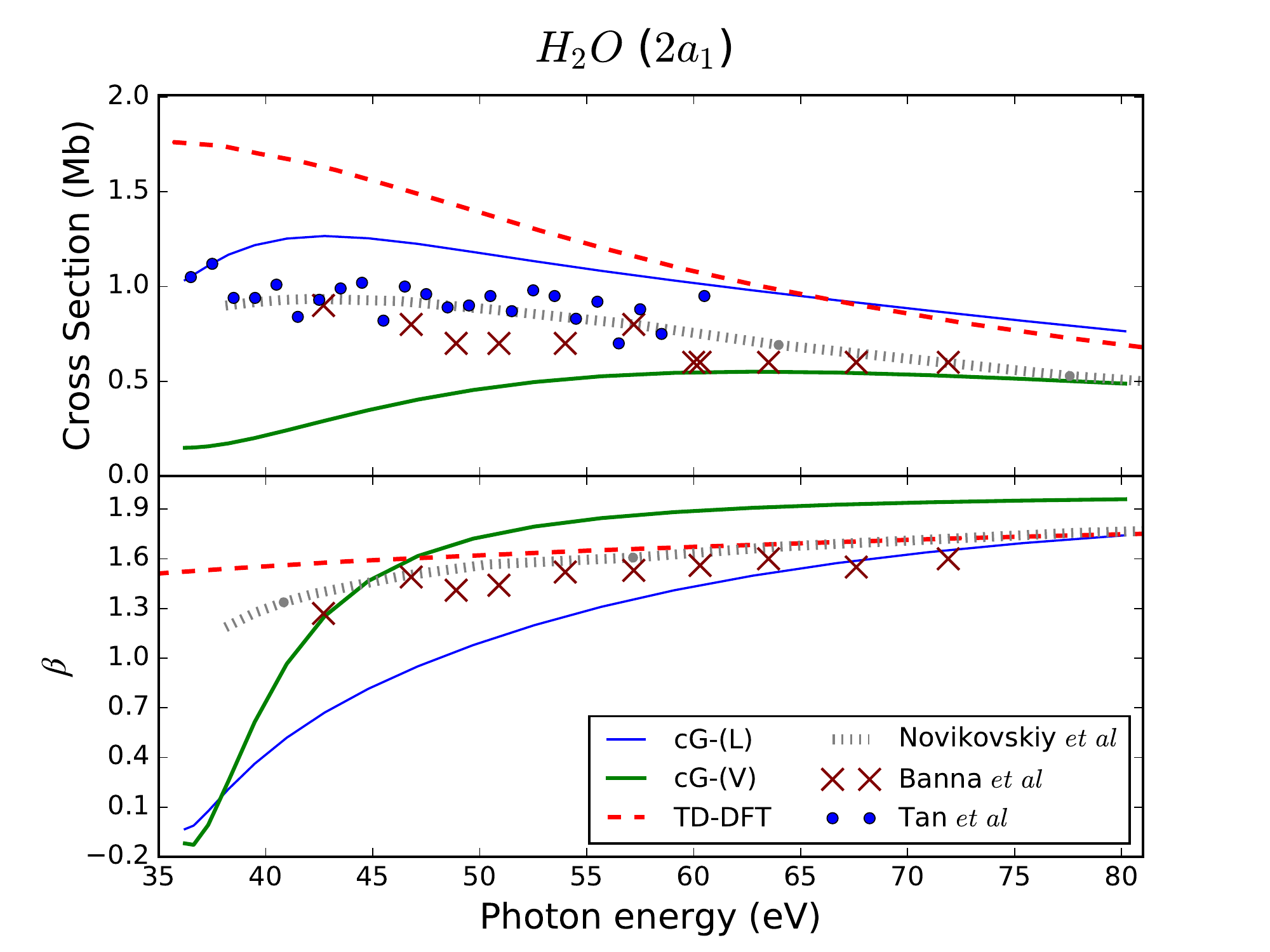}
\end{center}
\caption{Same as Fig. \ref{fig:H2Ocomp1b1} for  orbital $2a_1$ of H$_2$O.
}
\label{fig:H2Ocomp2a1}
\end{figure}

Results for orbitals $1b_1$, $3a_1$, $1b_2$ and $2a_1$ are shown in Figs. \ref{fig:H2Ocomp1b1}, \ref{fig:H2Ocomp3a1}, \ref{fig:H2Ocomp1b2} and \ref{fig:H2Ocomp2a1}, respectively, together with other theoretical results\cite{stener2002,novikovskiy2019,moitra2020} and experimental points.\cite{tan1978,truesdale1982,banna1986}
Note that once again experimental uncertainties in the literature are typically of the order of $5 \%$ (not shown in the figures).
The photoionization parameters for outer valence orbitals $1b_1$ and $3a_1$ are well reproduced. The velocity gauge results present an overall better agreement with experimental measurements (again, the length gauge results are shown here for fairness and completeness). Difficulties appears when looking at orbital $1b_2$, where the difference between the two gauges is more clearly visible and the results for $\beta(k_e)$ show larger deviations with respect to other calculations and experimental results. For orbital $2a_1$, the cGTO results are still reasonable, considering the disagreement observed between the two experimental and two theoretical sets of data plotted in the figure. 
The two general trends observed for ammonia are also featured for water: agreement for cross sections is generally better than for the more sensitive asymmetry parameter, and results are of relatively lower quality for inner versus outer orbitals.

\section{\sffamily \Large CONCLUSION \label{sec_conclusion}}

Studying molecular photoionization processes requires a good description of the outgoing electron.
At first sight, Gaussian representations do not seem very appropriate for highly oscillating wavefunctions. However we have shown that it is possible to use cGTO representations of the continuum functions in the context of photoionization calculations, while keeping the same level of accuracy for partial cross-sections and asymmetry parameters.
Formulas have been here derived within the framework of the monocentric approximation.
Transition integrals involving cGTOs have been written in closed form which can be easily evaluated at little numerical cost. The preliminary optimization giving optimal exponents for the Gaussian sets within a given energy range is costly but it had only to be performed once for each orbital quantum number. We have provided a table of such exponents that can be used for other applications involving continuum Coulomb or distorted states
with energy up to 45 eV;
the linear coefficients can be easily obtained through linear optimization.
Our complex Gaussian approach has been validated here by studying photoionization of two  molecules of type AH$_n$, namely ammonia and water.
Practically no difference is seen between a conventional calculation and a closed form approach through the proposed cGTO representation.

A similar conclusion may be reached by studying the photodetachment of anions. In this case, the Coulomb wave with charge $z=1$ are to be replaced by plane waves, and the cGTO exponents have to be calculated for a set of spherical Bessel functions of first kind instead of the regular Coulomb functions,\cite{ammar_phd} i.e. formula (\ref{eq:RegCoulFun}) with $z=0$.

The present investigation focused on a cGTO representation of the continuum states. For the initial bound state $\phi_i(\mathbf{r})$ we took a single center expansion over Slater-type orbitals. In the case where GTO were used instead, the involved matrix elements are also all analytical.
Actually, in such an all Gaussian approach, the radial integrals turn out to be even simpler because the integrands involves only Gaussian type exponentials and powers of the radial coordinate. The resulting closed form expressions have been presented in ref. \citenum{ammar2020}, and have been numerically validated with an rGTO representation of the exact states of the hydrogen atom \cite{ammar2020} and of approximate states of small atoms.\cite{ammar_phd}

We are currently working on the natural extension of the  proposed complex Gaussian strategy to represent continuum states in two directions.
The first one is to use cGTO to deal with other applications such as $(e,2e)$ molecular ionization under electron impact,
and the study of High Harmonic Generation.\cite{HHG1,HHG2,HHG3,HHG4}
The second direction, even more interesting, is related to the well-known Gaussian product theorem leading to drastic simplifications in the calculation of multicentric Gaussian integrals. This well-known property remains valid in the presence of complex exponents\cite{kuang1997_I,kuang1997_II}
and this motivates us to develop a generalization of the cGTO transition integrals, considered here in the monocentric approximation, to a more realistic multicentric model.
This should be made by taking into account the multicentric nature of the molecular target, in both initial (bound) and final (continuum) states.
In a preliminary investigation,\cite{ammar_phd,ammar2021}
we have already been able to show that the analytical character of the calculations, which is related to using cGTO representations of the continuum,
holds with a multicentric test-case initial state of Gaussian type. Whether the latter is described in spherical or cartesian coordinates, the matrix elements evaluation involves
more laborious mathematics.
Remains to apply the formulation to photoionization of a realistic molecular model.
As a second step, a similar multicentric modification in the final state should be considered.
This would ultimately drastically speed up numerical calculation for processes implying electronic continua in larger molecules.

\clearpage
\begin{table*}[ht]
{\scriptsize
\hspace{-1cm}
\begin{tabular}
{|c|c|c|c|c|c|}\hline
$i$ & $ \{\alpha_i\}_{l=0}    $ & $ \{\alpha_i\}_{l=1}    $ & $ \{\alpha_i\}_{l=2}    $ & $ \{\alpha_i\}_{l=3}    $ & $ \{\alpha_i\}_{l=4}    $ \\
\hline
1 & $ 0.00035555 + 0.00901614 \imath $ & $ 0.0001000 + 0.0151526 \imath $ & $ 0.00100000 + 0.00845442 \imath $ & $ 0.00014763 + 0.00587055 \imath $ & $ 0.00039019 - 0.01494676 \imath $ \\
2 & $ 0.00053211 + 0.02065728 \imath $ & $ 0.0001418 - 0.0171753 \imath $ & $ 0.00121886 + 0.01803494 \imath $ & $ 0.00109586 + 0.02784779 \imath $ & $ 0.00262171 - 0.03255562 \imath $ \\
3 & $ 0.00080009 - 0.01912061 \imath $ & $ 0.0013875 - 0.0021493 \imath $ & $ 0.00148684 + 0.01813637 \imath $ & $ 0.00148998 - 0.01168882 \imath $ & $ 0.00339622 + 0.01408910 \imath $ \\
4 & $ 0.00117085 - 0.01123799 \imath $ & $ 0.0019035 + 0.0216135 \imath $ & $ 0.00181381 + 0.02949728 \imath $ & $ 0.00205346 + 0.01943320 \imath $ & $ 0.00443045 - 0.02409112 \imath $ \\
5 & $ 0.00171276 + 0.02161259 \imath $ & $ 0.0025827 + 0.0298661 \imath $ & $ 0.00220490 + 0.02447032 \imath $ & $ 0.00280227 + 0.03734991 \imath $ & $ 0.00574632 + 0.02828016 \imath $ \\
6 & $ 0.00249700 + 0.03158584 \imath $ & $ 0.0034908 - 0.0241050 \imath $ & $ 0.00267806 - 0.01024824 \imath $ & $ 0.00358933 + 0.01238024 \imath $ & $ 0.00744649 + 0.02342353 \imath $ \\
7 & $ 0.00351122 - 0.03766107 \imath $ & $ 0.0046754 - 0.0338441 \imath $ & $ 0.00325391 - 0.01961417 \imath $ & $ 0.00459090 - 0.02321602 \imath $ & $ 0.00970895 + 0.03880653 \imath $ \\
8 & $ 0.00494586 - 0.02649727 \imath $ & $ 0.0061876 + 0.0378398 \imath $ & $ 0.00395324 - 0.02477460 \imath $ & $ 0.00588757 - 0.02219062 \imath $ & $ 0.01223182 + 0.02067056 \imath $ \\
9 & $ 0.00695348 - 0.02692584 \imath $ & $ 0.0156781 + 0.0504851 \imath $ & $ 0.00480634 - 0.01801510 \imath $ & $ 0.00757308 - 0.03698798 \imath $ & $ 0.01537803 - 0.03985771 \imath $ \\
10 & $ 0.00974499 - 0.04400129 \imath $ & $ 0.0196315 - 0.0395359 \imath $ & $ 0.00585038 - 0.02149702 \imath $ & $ 0.00967950 - 0.02915329 \imath $ & $ 0.01953215 - 0.03489981 \imath $ \\
11 & $ 0.01341500 + 0.03611724 \imath $ & $ 0.0246020 - 0.0352098 \imath $ & $ 0.00712456 - 0.02735801 \imath $ & $ 0.01239829 + 0.04894840 \imath $ & $ 0.02499666 + 0.05443031 \imath $ \\
12 & $ 0.01853987 + 0.04014546 \imath $ & $ 0.0308382 - 0.0302620 \imath $ & $ 0.00867609 - 0.03810200 \imath $ & $ 0.01699053 - 0.05180981 \imath $ & $ 0.03507324 - 0.05430569 \imath $ \\
13 & $ 0.02554489 - 0.05549019 \imath $ & $ 0.0386508 - 0.0436192 \imath $ & $ 0.01058491 + 0.03742598 \imath $ & $ 0.03292671 + 0.06650370 \imath $ & $ 0.04435803 - 0.02396048 \imath $ \\
14 & $ 0.03437502 + 0.04391841 \imath $ & $ 0.0484105 + 0.0624525 \imath $ & $ 0.01305188 + 0.03757777 \imath $ & $ 0.04217102 - 0.07408910 \imath $ & $ 0.05617605 + 0.02052618 \imath $ \\
15 & $ 0.04626602 + 0.05104620 \imath $ & $ 0.0618203 + 0.0513264 \imath $ & $ 0.01646720 - 0.05003223 \imath $ & $ 0.06196807 + 0.03814310 \imath $ & $ 0.07128007 + 0.02106443 \imath $ \\
16 & $ 0.06218273 - 0.06475675 \imath $ & $ 0.0816826 - 0.0552154 \imath $ & $ 0.02076651 + 0.05258585 \imath $ & $ 0.19225626 + 0.03964722 \imath $ & $ 0.16136565 + 0.02940124 \imath $ \\
17 & $ 0.08554072 - 0.02426909 \imath $ & $ 0.1936246 + 0.0097332 \imath $ & $ 0.03514922 - 0.06965770 \imath $ & $ 0.30147472 + 0.00811515 \imath $ & $ 0.39894781 + 0.02631795 \imath $ \\
18 & $ 0.11861245 + 0.07471136 \imath $ & $ 0.2915597 + 0.0024838 \imath $ & $ 0.04424413 + 0.07302765 \imath $ & $ 0.51617609 + 0.04399335 \imath $ & $ 0.55207468 + 0.07285449 \imath $ \\
19 & $ 0.16691697 + 0.08944899 \imath $ & $ 0.4971406 - 0.0113366 \imath $ & $ 0.09707468 - 0.10000000 \imath $ & $ 0.82002155 + 0.05919349 \imath $ & $ 0.86945734 - 0.01791204 \imath $ \\
20 & $ 0.23869247 - 0.09999999 \imath $ & $ 0.8532849 + 0.0200975 \imath $ & $ 0.12043971 + 0.08910818 \imath $ & $ 1.19395577 + 0.01279526 \imath $ & $ 1.12048249 + 0.08602790 \imath $ \\
21 & $ 1.19614463 - 0.02970423 \imath $ & $ 1.3797714 + 0.0062079 \imath $ & $ 2.10573048 - 0.03921720 \imath $ & $ 1.48783709 + 0.02469442 \imath $ & $ 1.41353953 + 0.04522398 \imath $ \\
22 & $ 5.33188880 - 0.01441299 \imath $ & $ 2.1958583 - 0.0311758 \imath $ & $ 3.94194435 + 0.02888234 \imath $ & $ 2.19692967 - 0.08785385 \imath $ & $ 2.14872735 - 0.01352875 \imath $ \\
23 & $ 7.71182758 - 0.09811990 \imath $ & $ 3.5672935 - 0.0087858 \imath $ & $ 6.10200559 - 0.01548483 \imath $ & $ 3.54466390 + 0.02398460 \imath $ & $ 3.60699097 - 0.03483872 \imath $ \\
24 & $ 10.9962323 - 0.09305650 \imath $ & $ 5.7419028 + 0.0012293 \imath $ & $ 9.14648581 - 0.00307356 \imath $ & $ 5.54622769 + 0.00914500 \imath $ & $ 5.64732513 + 0.01125528 \imath $ \\
25 & $ 15.5241861 - 0.09730549 \imath $ & $ 9.2535835 - 0.0212019 \imath $ & $ 13.7297058 - 0.03200391 \imath $ & $ 9.24463017 - 0.00535522 \imath $ & $ 9.19300771 + 0.04174570 \imath $ \\
26 & $ 21.7519013 - 0.09157168 \imath $ & $ 14.868739 - 0.0267870 \imath $ & $ 20.4351559 - 0.00138094 \imath $ & $ 14.8708853 - 0.02233793 \imath $ & $ 14.8869767 - 0.03464065 \imath $ \\
27 & $ 30.2989535 - 0.07925670 \imath $ & $ 23.954872 - 0.0406721 \imath $ & $ 30.3651118 + 0.02650005 \imath $ & $ 23.9508076 + 0.08258257 \imath $ & $ 23.9066166 + 0.00726798 \imath $ \\
28 & $ 42.0163884 - 0.09927635 \imath $ & $ 38.575864 + 0.0168896 \imath $ & $ 45.2156191 - 0.00722570 \imath $ & $ 38.5477254 + 0.09948084 \imath $ & $ 38.5167312 - 0.01803004 \imath $ \\
29 & $ 58.1483705 - 0.08991001 \imath $ & $ 62.088037 - 0.0059998 \imath $ & $ 67.2295184 + 0.01284461 \imath $ & $ 62.0907276 - 0.03652093 \imath $ & $ 62.0736229 - 0.06204804 \imath $ \\
30 & $ 80.7346781 - 0.08528403 \imath $ & $ 99.986651 + 0.0164035 \imath $ & $ 99.9856251 - 0.06227840 \imath $ & $ 100.021625 + 0.00412096 \imath $ & $ 99.9767615 - 0.03846239 \imath $ \\
\hline 					
\end{tabular}
}
\caption{Optimal cGTO exponents $\{\alpha_i\}_{l}$ obtained after fitting the sets of Coulomb functions  as defined in eq. \eqref{eq:ensembleF}, for $l=0,1,2,3,4$. The optimal exponents for $l=1$ have been already published in ref. \citenum{ammar2020}
and were obtained with slightly different numerical parameters in the optimization code. \label{tab:exponents}}
\end{table*}

\subsection*{\sffamily \large ACKNOWLEDGMENTS}

We acknowledge Torsha Moitra, Sonia Coriani and Nikolay M. Novikovskiy for sending us their raw data associated with refs \citenum{moitra2020} and \citenum{novikovskiy2019}.
The PMMS (P\^ole Messin de Mod\'elisation et de Simulation) is also acknowledged for providing us with computer time.

\subsection*{\sffamily \large DATA AVAILABILITY STATEMENT}

The data that support the findings of this study are available from the corresponding author upon reasonable request.






\end{document}